\documentclass{aa}
\usepackage{graphicx}

\begin{document}
\title{A library of high resolution synthetic stellar spectra from 300nm to
1.8${\rm \mu}$m with solar and $\alpha$-enhanced composition}
\author{P. Coelho\inst{1,2}
\and
 B. Barbuy\inst{1}
\and
J. Mel\'endez\inst{3}
\and
 R. P. Schiavon\inst{4}
\and
 B. V. Castilho\inst{5}
 }
\offprints{P. Coelho}
\institute{
Universidade de S\~ao Paulo, Rua do Mat\~ao 1226, S\~ao Paulo 05508-900, 
Brazil; pcoelho@usp.br, barbuy@astro.iag.usp.br.
\and
Max-Planck-Institut f\"ur
 Astrophysik, Karl-Schwarzschild-Strasse 1, D-85741 Garching, Germany;  pcoelho@mpa-garching.mpg.de
\and
Department of Astronomy, Caltech, 1200 E. California Blvd,
Pasadena, CA 91125; jorge@astro.caltech.edu
\and
Department of Astronomy, University of Virginia, P.O. Box 3818,
Charlottesville, VA 22903-0818, USA; 
ripisc@virginia.edu
\and
Laborat\'orio Nacional de Astrof\'{\i}sica / MCT, CP 21, 37500-000
Itajub\'a, Brazil; bruno@lna.br
}

\date{}

\abstract{
Libraries of stellar spectra are fundamental tools for the study
of stellar populations and both
empirical and synthetic libraries have been used for this purpose. 
In this paper, a new library of high resolution synthetic spectra is presented,
ranging from the near-ultraviolet (300nm) to the near-infrared
(1.8${\rm \mu}$m). The library spans all the stellar types that are 
relevant to the integrated light of old and intermediate-age
stellar populations in the involved spectral region (spectral types F
through M and all luminosity classes). The grid was computed for 
metallicities ranging from [Fe/H] = --2.5 to +0.5, including both solar
and $\alpha$-enhanced ([$\alpha$/Fe] = 0.4) chemical compositions.
The synthetic spectra are a good match to observations of stars
throughout the stellar parameter space encompassed by the library and
over the whole spectral region covered by the computations. 
\keywords{Stars: atmospheres, Stars: spectra, Stars: population synthesis} }

\titlerunning{A library of synthetic spectra with $\alpha$-enhancement}{} 
\authorrunning{P. Coelho et al.}

\maketitle

\section{Introduction} 

Libraries of stellar spectra are one of the main ingredients of stellar
population synthesis models (e.g.  Bruzual \& Charlot 1993, 2003; Cervi\~no
\& Mas-Hesse 1994; Fioc \& Rocca-Volmerange 1997; Leitherer
et al. 1999; Vazdekis 1999; Buzzoni 2002; Schulz et al. 2002; Jimenez
et al. 2004; Gonzalez Delgado et al. 2005; Maraston 2005; Schiavon 2005) and 
both empirical and theoretical libraries 
have improved dramatically in recent years. The first
empirical libraries (e.g. Gunn \& Stryker 1983; Jacoby et al. 1984)
used in stellar population synthesis work were restricted to 
a relatively small number of stars with uncertain atmospheric
parameters.  The quality of empirical libraries has been refined along the years (e.g. 
Pickles 1998; Jones 1999) and recently, a major improvement 
has been achieved by Le Borgne et al. (2003; STELIB) and Valdes et al. (2004;
Indo-US), which provide high S/N, medium resolution (down to FWHM $\sim$
1${\rm \AA}$) and a good coverage of the color-magnitude diagram.

Amongst the synthetic libraries, perhaps the most widely
used is the flux distribution predicted from the Kurucz (1993)
model atmospheres. Lejeune et al. (1997, 1998) and Westera et al. (2002)
extended this library to include spectra of M stars computed
from model atmospheres from Fluks et al. (1994), Bessell et al. (1989), 
Bessell (1991) and Allard \& Hauschildt (1995).  Moreover, they calibrated the flux distribution of the
synthetic spectra in order to match the colors of observed stars (BaSeL
library). However, the spectral resolution of the BaSeL
library is limited to the sampling of the model atmosphere grid ($\sim$
20 ${\rm \AA}$) which is by far lower than the modern observed spectra
of both individual stars and integrated stellar populations.
More recently, resolution ceased to be a limitation, with
the publication of high resolution spectral libraries by Chavez et al. (1997), Murphy \&
Meiksin (2004), Martins et al. (2005) and Rodrigues-Merino et al (2005). However, these 
theoretical libraries still
have a more limited wavelength coverage than that of the previous
low-resolution synthetic libraries.

The choice of using either an empirical or a synthetic library in stellar 
population models is a subject of debate in the literature. One disadvantage of synthetic 
libraries is that they rely on model atmospheres, which are subject to 
systematic uncertainties. Besides, computing a reliable 
high-resolution synthetic spectral library for a large range of stellar 
parameters and in a wide spectral region is a very challenging task, because it 
requires building an extensive and reliable list of atomic and molecular line 
opacities which are needed for an accurate reproduction of high-resolution 
spectra of real stars. On the other hand, synthetic 
spectral libraries overcome limitations of empirical libraries, the most 
important being their inability to extrapolate to abundance patterns 
that differ from that of the library stars, which are from the solar neighbourhood 
or, in some cases, that of
the Magellanic Clouds.
Therefore, models based on empirical libraries
cannot reproduce the integrated spectra of systems that have undergone
star formation histories that are different from that of local systems.

The first compelling evidence that models based on empirical libraries
have difficulty reproducing the integrated light of extra-galactic
populations was presented by Worthey et al. (1992). These authors 
showed that single stellar population
synthesis models for the Lick/IDS indices (Burstein et al. 1984; Worthey
et al. 1994; Trager et al. 1998) cannot reproduce the indices
measured in giant elliptical galaxies, thus indicating 
that these systems are overabundant in $\alpha$-elements
relative to the Sun. This happens because, by construction, the
abundance pattern of models based on empirical libraries is dictated by
that of the library stars, which mirrors the abundance pattern of the
solar neighborhood (but see discussion in Schiavon 2005). Therefore,
such models are bound to follow the relation between [Fe/H] and
[$\alpha$/Fe] that is characteristic of the solar neighborhood where
[$\alpha$/Fe] is above solar at low metallicities and around solar
in the high-metallicity regime (e.g. McWilliam 1997). As
a result, the models cannot match the integrated light of metal-rich,
$\alpha$-enhanced systems, such as giant ellipticals.

This deficiency could be in principle partially cured empirically, by
including spectra of bulge stars in the spectral libraries.
The population of the bulge of the Galaxy has an $\alpha$-enhanced
abundance pattern even at high [Fe/H] 
(e.g. McWilliam \& Rich 1994, 2004; Zoccali et al. 2004). However, their inclusion in 
sufficient numbers is hampered by their faintness and severe reddening, so that
very few (or none) of these stars are included in empirical libraries.

For the case of the Lick/IDS indices, models accounting for variable
$\alpha$-enhancement have been developed (e.g. Trager et al. 2000;  
Thomas et al. 2003a; Proctor et al. 2004; Tantalo et
al. 2004; Mendes de Oliveira et al. 2005; Schiavon 2005). Even though the Lick/IDS indices 
have already proven their critical importance for the understanding of the
stellar populations, it is crucial to develop models that can explore the huge amount 
of information contained in the whole spectrum, since methods fitting the entire 
spectrum are now
feasible (Heavens et al. 2004; Panter et al. 2004).

Currently, the only way of producing models with abundance patterns
that differ from that of the stars in the solar neighbourhood 
is through the adoption of synthetic stellar
spectra. Although grids of high-resolution synthetic spectra have
recently been computed, a wide grid with an $\alpha$-enhanced
mixture is still lacking. To our knowledge, so far two grids
of synthetic spectra with $\alpha$-enhanced mixtures were
available: Barbuy
et al. (2003) published a grid in the wavelength range 4600 - 5600 \AA\ aiming
at the study of the  effect of $\alpha$-enhancement on
the Lick/IDS indices Mg2, Fe5270 and Fe5335; and Zwitter et al. (2004)
published a grid covering 7653 - 8747 ${\rm \AA}$, which is aimed
to be used with the spectroscopic surveys by RAVE and GAIA (Steinmetz 2003; Katz 2004).

In this work we present for the first time a grid of
synthetic spectra covering solar and $\alpha$-enhanced
mixtures in a wide baseline. The grid is aimed at applications in
stellar population synthesis, specifically to the study of old and
intermediate-aged stellar populations. Thus, the grid covers
the effective temperatures and surface gravities suitable for these systems. 

We would like to emphasize, however, that applications of
this grid are not limited to stellar population studies.
The spectral library should also be a very useful tool for
atmospheric parameter determination work, in particular as input in 
automatic procedures of deriving atmospheric parameters
of a large number of observed stars (e.g. Cayrel et al. 1991a; Franchini et al.
2004; Willemsen et al. 2005; Girard \& Soubiran 2005; Valenti \& 
Fischer 2005).

This paper is organized as follows. 
In Section 2 we briefly describe the computation of the synthetic
spectra and the production of the atomic and molecular line 
lists. In Section 3 the library is presented and aspects of the relative
contribution of the chemical species are investigated. A Summary is
given in Section 4.

\section{Calculations}

The spectra were computed using the code \textit{PFANT}. The first version of this
code named \textit{FANTOM} was developed by Spite (1967) for the
calculation of atomic lines. Barbuy (1982)
included the calculation of molecular lines, implementing the dissociative
equilibrium by Tsuji (1973) and molecular line computations as described
in Cayrel et al. (1991b). 
It has been further improved for
 calculations of large wavelength coverage and inclusion of Hydrogen lines
as described in Barbuy et al. (2003). Ten hydrogen lines 
from $H_{10}$ to $H_{\alpha}$ are considered, through a revised version of
the code presented in Praderie (1967). Higher order Balmer lines and Paschen 
and Brackett series are not yet considered in the present version of the code.
Given a stellar model
atmosphere and lists of atomic and molecular lines, the code computes 
a synthetic spectrum assuming local thermodynamic equilibrium. 

\begin{figure*}
\includegraphics[height=18cm,angle=270]{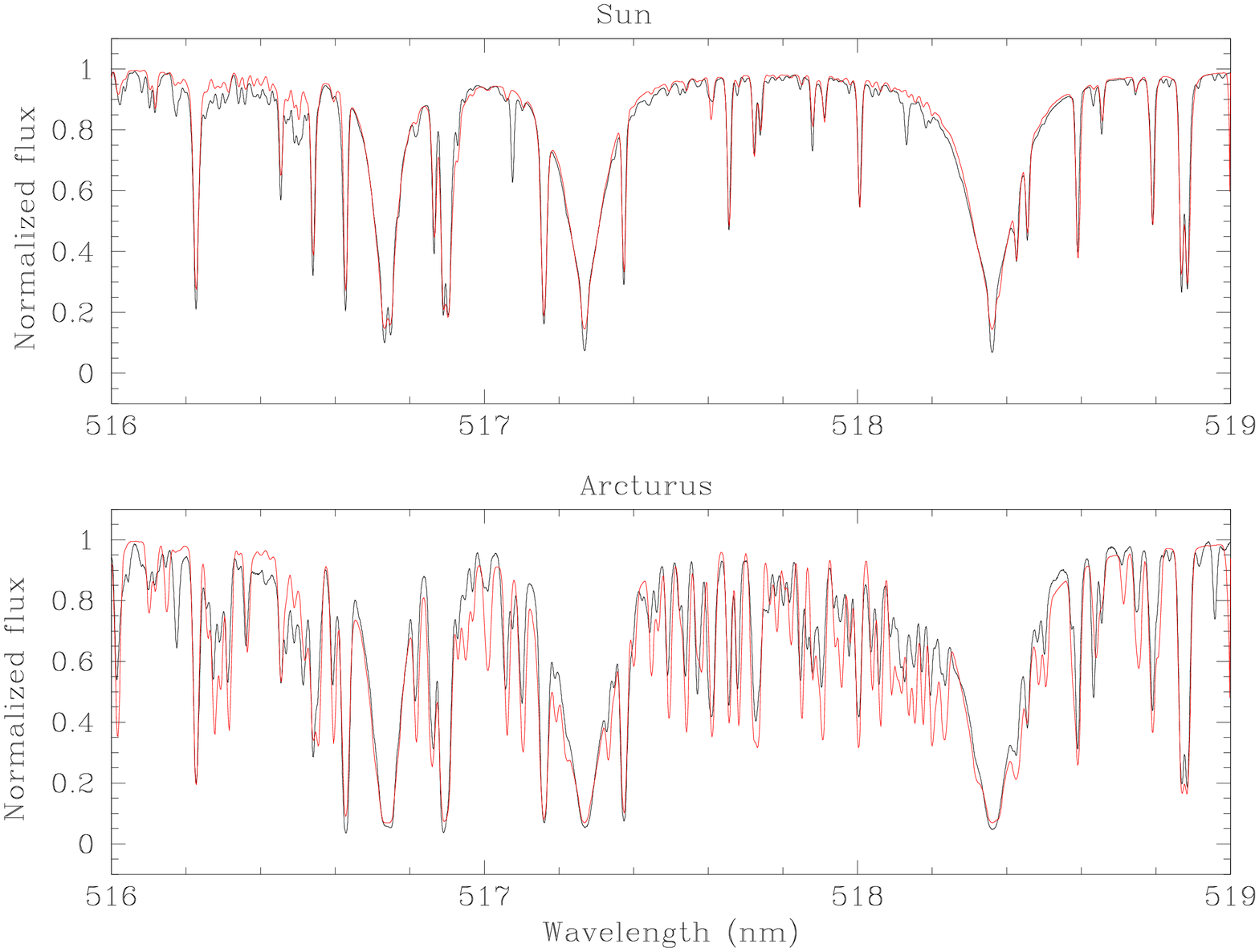} 
\caption{
\textit{Upper panel:} comparison of the observed solar spectrum (thin line) and 
synthetic spectrum (thick line). \textit{Lower panel:} comparison of 
Arcturus observed spectrum (thin line) and synthetic spectrum (thick line). 
Adopted parameters are given in the text.}
\label{fig_compac1}
\end{figure*}

\begin{figure*}
\includegraphics[height=18cm,angle=270]{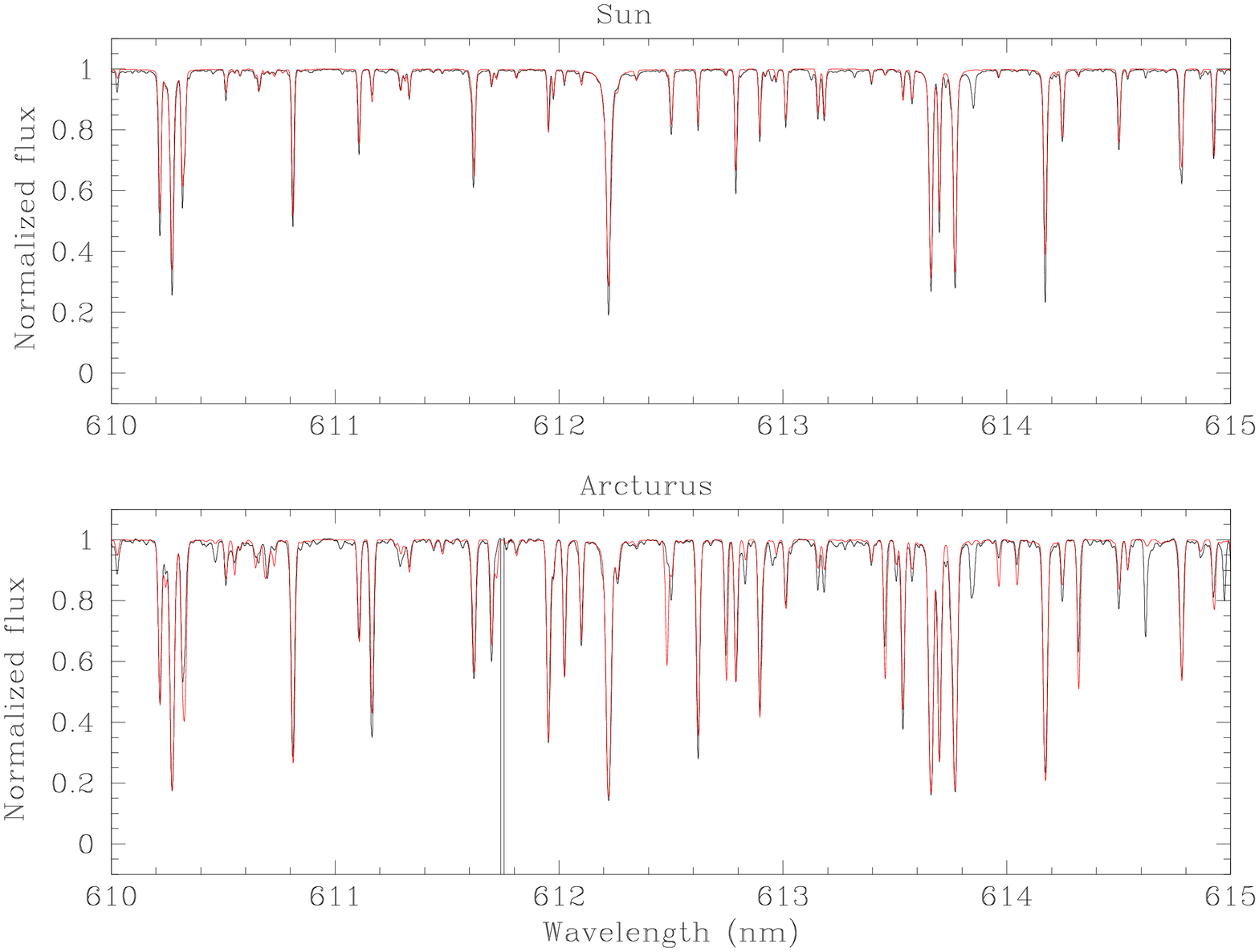} 
\caption{Same as Fig. 1 for the region 610-615 nm.}
\label{fig_compac2}
\end{figure*}

The model atmospheres adopted were those presented by Castelli
\& Kurucz (2003), which
account for $\alpha$-enhanced compositions. The solar
abundances adopted are those from Grevesse \& Sauval (1998), which
are consistent with the abundances used in the model atmospheres.
These models differ from the previous ATLAS9 grid (Kurucz 1993) mainly
because they include new opacity distribution functions, new molecular
opacities and do not include overshooting.

These models have mixing length to pressure scale height ratio
of 2.0. A lower value ($l/H_P = 0.5$) was suggested by Fuhrmann et al.
(1993) and Van 't Veer-Menneret \& M\'egessier (1996) as being more 
suitable to reproduce the profile of Balmer lines (as was adopted in the models 
in Barbuy et al. 2003). But it is important to note that the Balmer
lines are not well reproduced also because
their cores are affected by the chromosphere
and by departures from local thermodynamic equilibrium.

\subsection{Atomic line opacities}

Accurate line opacities are in the very heart of a successful synthetic
spectrum computation. 
The line list employed is an updated version of the one presented
in Barbuy et al. (2003), where oscillator strengths (\textit{log
gf}s) and damping constants were revised. 

The line list was extended up
to 1.8${\rm \mu}$m with the inclusion of the data from Mel\'endez
\& Barbuy (1999). The most recent oscillator strengths listed in
the latest electronic version of the NIST database (Reader et al. 2002)
were adopted. These values account for 10 to 50\% of the lines in
the near-UV and visible spectral regions, respectively.
The \textit{log gf} values of 139 Fe II lines were updated according to the
normalization given in Mel\'endez \& Barbuy (2002). Line by line
fits to the solar spectrum were done in specific spectral
regions, and in this process $\sim$ 240 lines had their \textit{log
gf}s selected among values from Sneden et al. (1996), Kupka et al. (1999; VALD),
Biehl (1976), Steffen (1985), McWilliam \& Rich
(1994) and Bensby et al. (2003). Additionally, 21 hyperfine structures
for the heavy elements Ba, La, Eu, Co, Pb were included, adopting 
a solar isotopic mix. For the hyperfine structures, 
a code described and made available by McWilliam (1998) is employed,
as described in Allen (2005).  

Accurate values for the damping constant $\gamma$ are also important
for the computation of line profiles. This is crucial
for the strong lines, and it also has a non-negligible effect
on weaker lines (see Ryan 1998 for a discussion). 

In our atomic line list, 36\% of the lines had the collisional 
broadening obtained from Anstee \& O'Mara (1995), Barklem \& O'Mara 
(1997), Barklem, O'Mara \& Ross (1998) and Barklem et al. 2000 
(this series of papers is referred to hereafter as ABO). Those 
lines correspond to the totality of the strong neutral lines.  The other 
lines were either of higher ionization stages 
\footnote{Cross sections for ionized lines are not easily computed
(see Barklem \& O'Mara 1998 and Barklem et al. 2000).
The cross sections for ionized transitions presented in Barklem et al. (2000) 
were included in our line list.}
or were not in the energy domain range of 
the tabulated cross sections. Thus these lines were either manually fitted to 
the Sun or were assumed to have interaction constant
$C_6 = 0.3 \times 10^{-31}$.

Around 5\% of the lines which had $C_6$ derived from ABO cross sections 
had to be manually fitted in order to reproduce the solar spectrum. 
This happened due to the dependence of the line profile both on the collisional 
broadening and on the 
model atmosphere employed. The conversion of the ABO cross sections into $C_6$
values are detailed 
in Appendix A. 

The high quality of the final line list can be assessed by comparisons
with high resolution spectra of the Sun (Kurucz et al. 1984) and
Arcturus (Hinkle et al. 2000). The resolution of these spectra are of
the order R $\sim$ $10^5$, thus allowing a thorough verification
of the accuracy of our wavelengths, oscillator strengths and damping
constants. In Figures \ref{fig_compac1} and \ref{fig_compac2}, 
comparisons of our synthetic spectra to those of the Sun and Arcturus
are shown in two spectral regions. For the Sun  the synthetic
spectrum was computed adopting T$_{\rm eff}$ = 5770K, log g = 4.44,
microturbulent velocity v$_{\rm t}$ = 1.0 km/s and abundances from Grevesse \&
Sauval (1998). For Arcturus the parameters adopted were T$_{\rm
eff}$ = 4275 K, log g = 1.55 and v$_{\rm t}$ = 1.65 km/s (Mel\'endez
et al. 2003). An ATLAS9 model atmosphere
with these parameters was kindly computed by
F. Castelli (private comunication). For the abundances, we adopted [Fe/H] = -0.54,
[C/Fe] = -0.08, [N/Fe] = +0.30, [O/Fe] = 0.43 and [Ni/Fe] = +0.02 from
 Mel\'endez et al. (2003), [Mg/Fe] = +0.30 and [Ca/Fe] =
+0.17 from McWilliam \& Rich (1994).  

In both figures the Sun is very well reproduced. For the case of Arcturus,
the synthetic spectrum looks somewhat strong-lined. At such a
high resolution, small errors in the atmospheric parameters can easily
account for this difference, and the overall agreement is very good.

When comparing synthetic spectrum against observed spectrum, 
one also has to be aware that observed spectra taken through the atmosphere
include telluric lines, whereas computed spectra do not. A crucial aspect of our procedure is the
derivation of reliable oscillator strengths from comparison with ground-based 
observations of the Sun and Arcturus, as explained in Barbuy et al. (2003). This 
however is not feasible in regions of the spectra which are heavily contaminated by telluric lines.
For that reason, there is a gap in the line list between 1.35 and 1.50 ${\rm \mu}$m 
- where the telluric absorption is severe - since we refrained from adopting 
lines with purely theoretical \textit{log gf}s in that region.

\subsection{Molecular line opacities}

\begin{figure}   
\includegraphics[height=8.8cm,angle=270]{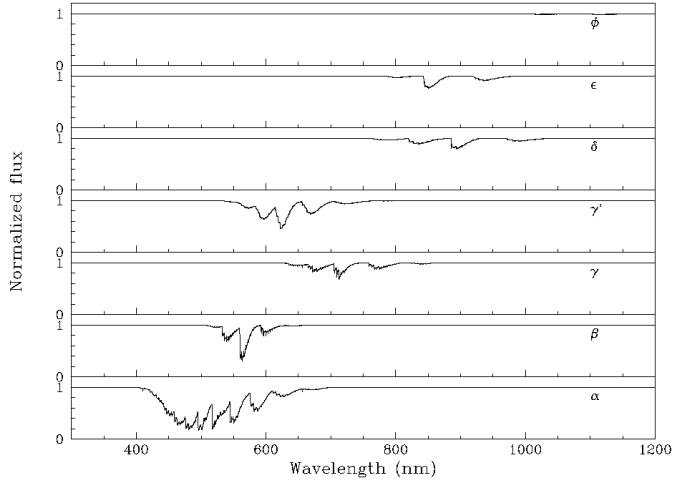} 
\caption{Each of the seven TiO systems considered in this work, computed 
separately. The identification of each band is given in the figure. These 
spectra were computed with (T$_{\rm eff}$, log g, [Fe/H], [$\alpha$/Fe]) = 
(3500, 0.0, 0.0, 0.0). }
\label{fig_tio0}
\end{figure}

Molecular lines are an important source of opacity in the
atmospheres of late-type stars. Molecular bands in fact dictate the
overall shape of the spectra of late-K and M stars, which are the dominant
contributors to the integrated light of old and intermediate-age stellar
populations longward of $\sim$ 650 nm.  Therefore, it is crucial to
model their intensities correctly if one wants to make accurate predictions
in these spectral regions. The molecules considered in the calculation
of the present library are listed in Table \ref{tab_mol}.

\begin{figure*}
\centerline{
\includegraphics[width=10.5cm,angle=270]{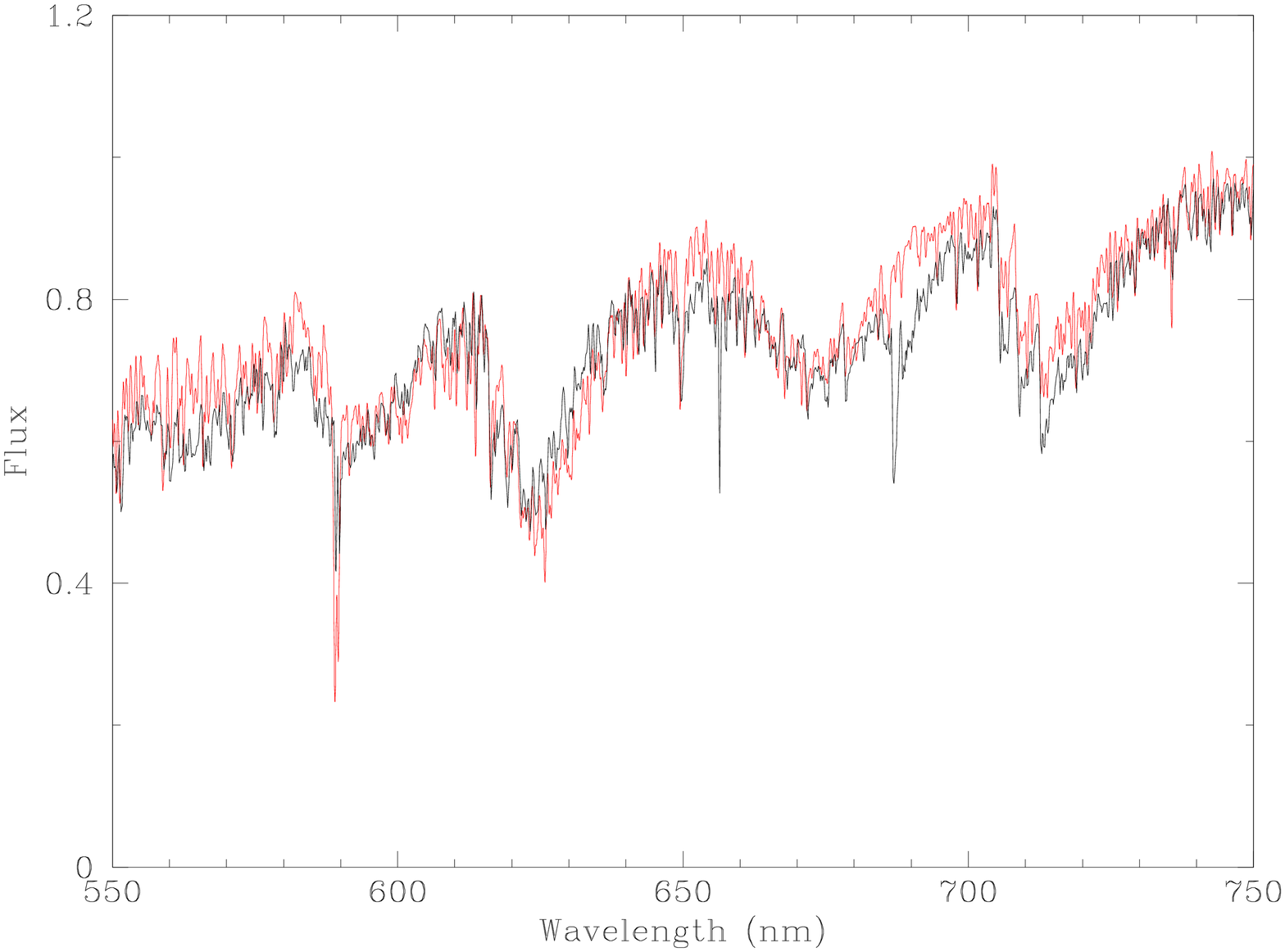}} 
\caption{Comparison between the star HD 102212 (thin line), atmospheric 
parameters (T$_{\rm eff}$, log g, [Fe/H]) = (3660, 0.0, 0.0), obtained from 
STELIB library and a synthetic spectrum (thick line), parameters (T$_{\rm eff}$, 
log g, [Fe/H], [$\alpha$/Fe]) = (3600, 0.0, 0.0, 0.0). This wavelength range 
illustrates the TiO $\gamma$ and $\gamma$' systems.}\label{fig_tio_1}
\end{figure*}
\begin{figure*}
\centerline{
\includegraphics[width=10.5cm,angle=270]{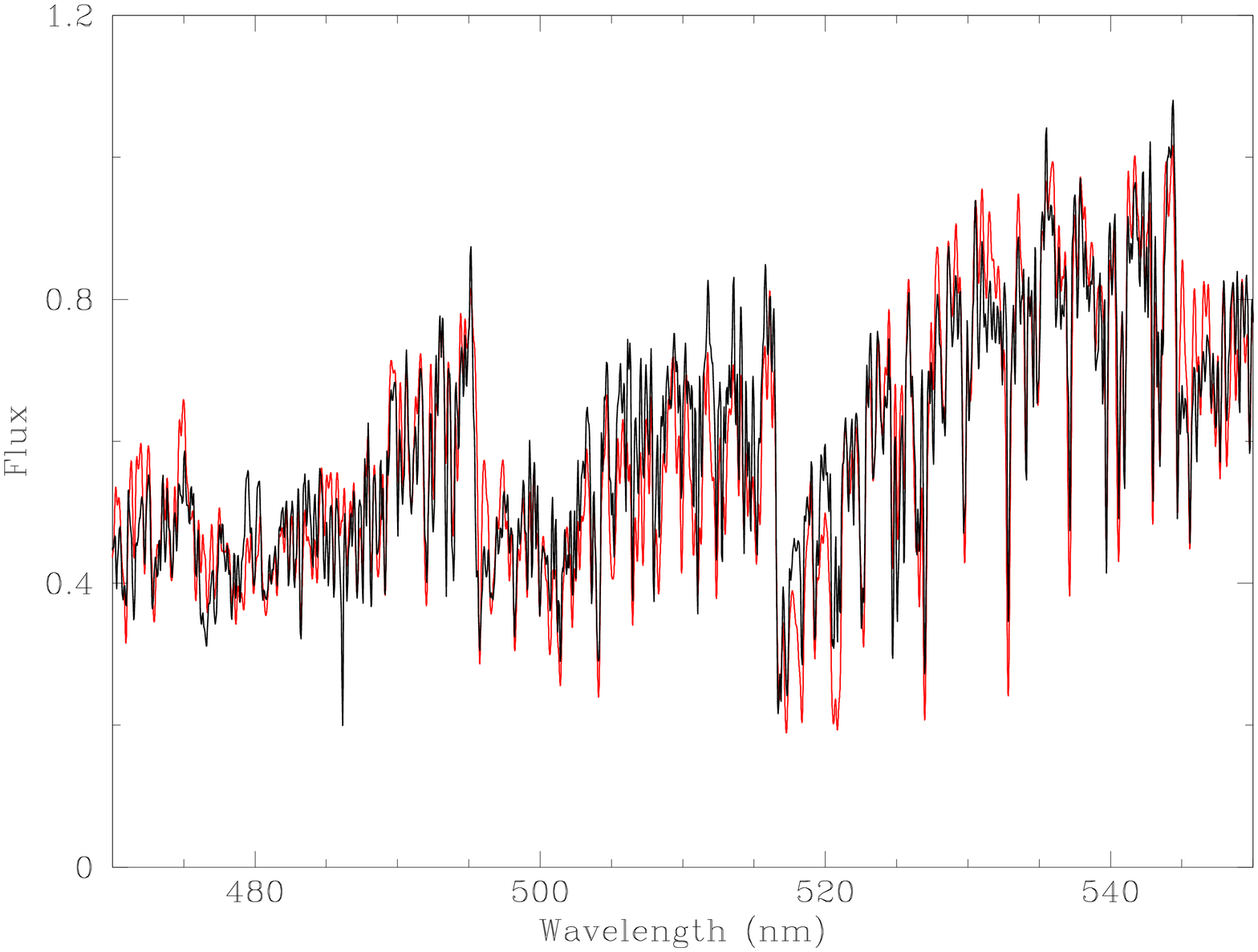} }
\caption{Comparison between the star HD 39801 (thin line), atmospheric 
parameters (T$_{\rm eff}$, log g, [Fe/H]) = (3540, 1.0, 0.0), obtained from 
Indo-US library and a synthetic spectrum (thick line), parameters (T$_{\rm 
eff}$, log g, [Fe/H], [$\alpha$/Fe]) = (3600, 1.0, 0.0, 0.0). This wavelength 
range illustrates the TiO $\alpha$ system.}\label{fig_tio_2}
\end{figure*}
\begin{figure*}
\centerline{
\includegraphics[width=10.5cm,angle=270]{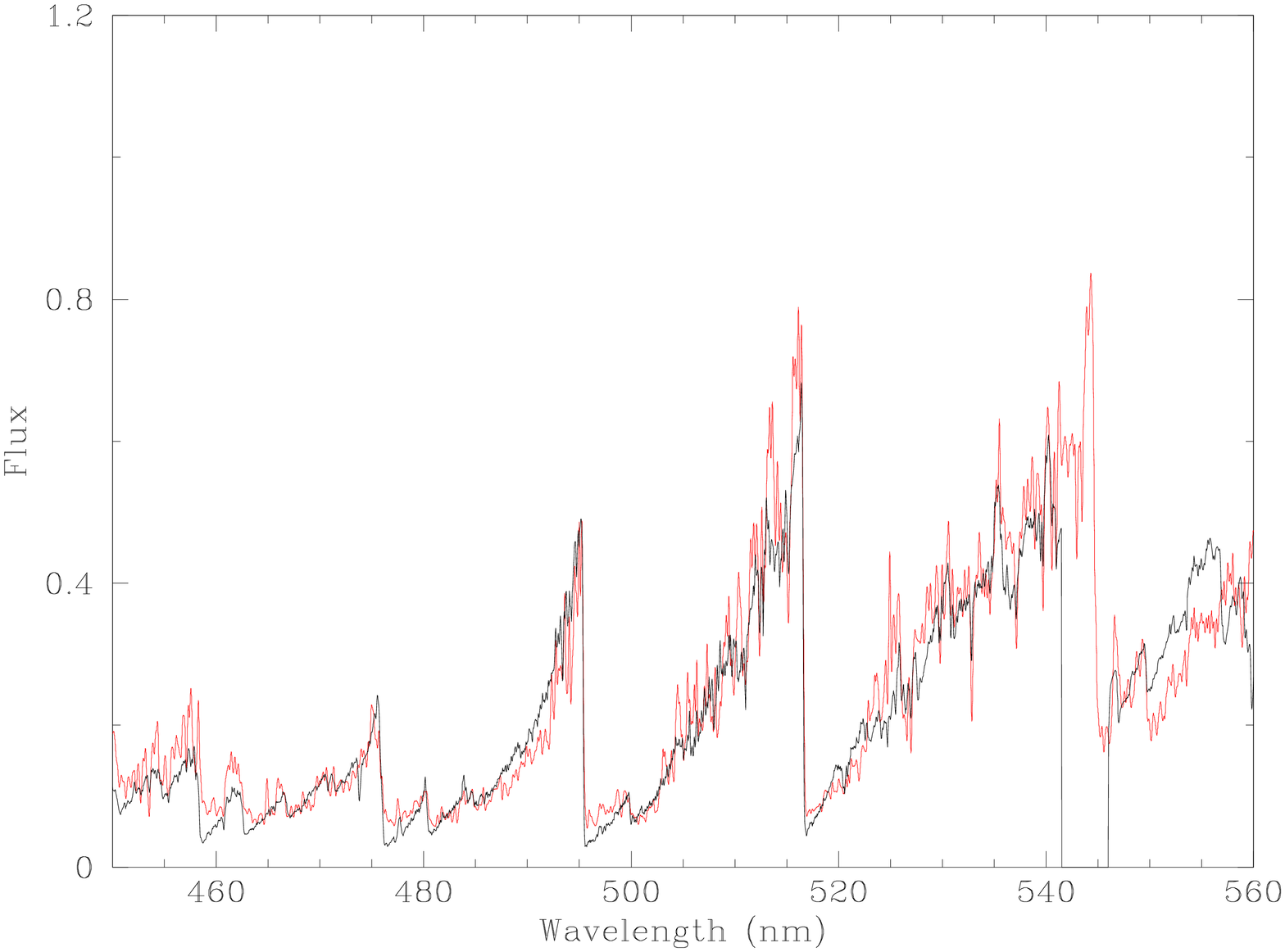} }
\caption{Comparison between the star HD 126327 (thin line), atmospheric 
parameters (T$_{\rm eff}$, log g, [Fe/H]) = (3000, 0.0, -0.58), obtained from 
Indo-US library and a synthetic spectrum (thick line), parameters (T$_{\rm eff}$, 
log g, [Fe/H], [$\alpha$/Fe]) = (3000, 0.0, 0.0, 0.0). This wavelength range 
illustrates the TiO $\alpha$ system.}\label{fig_tio_3}
\end{figure*}
\begin{figure*}
\centerline{
\includegraphics[width=10.5cm,angle=270]{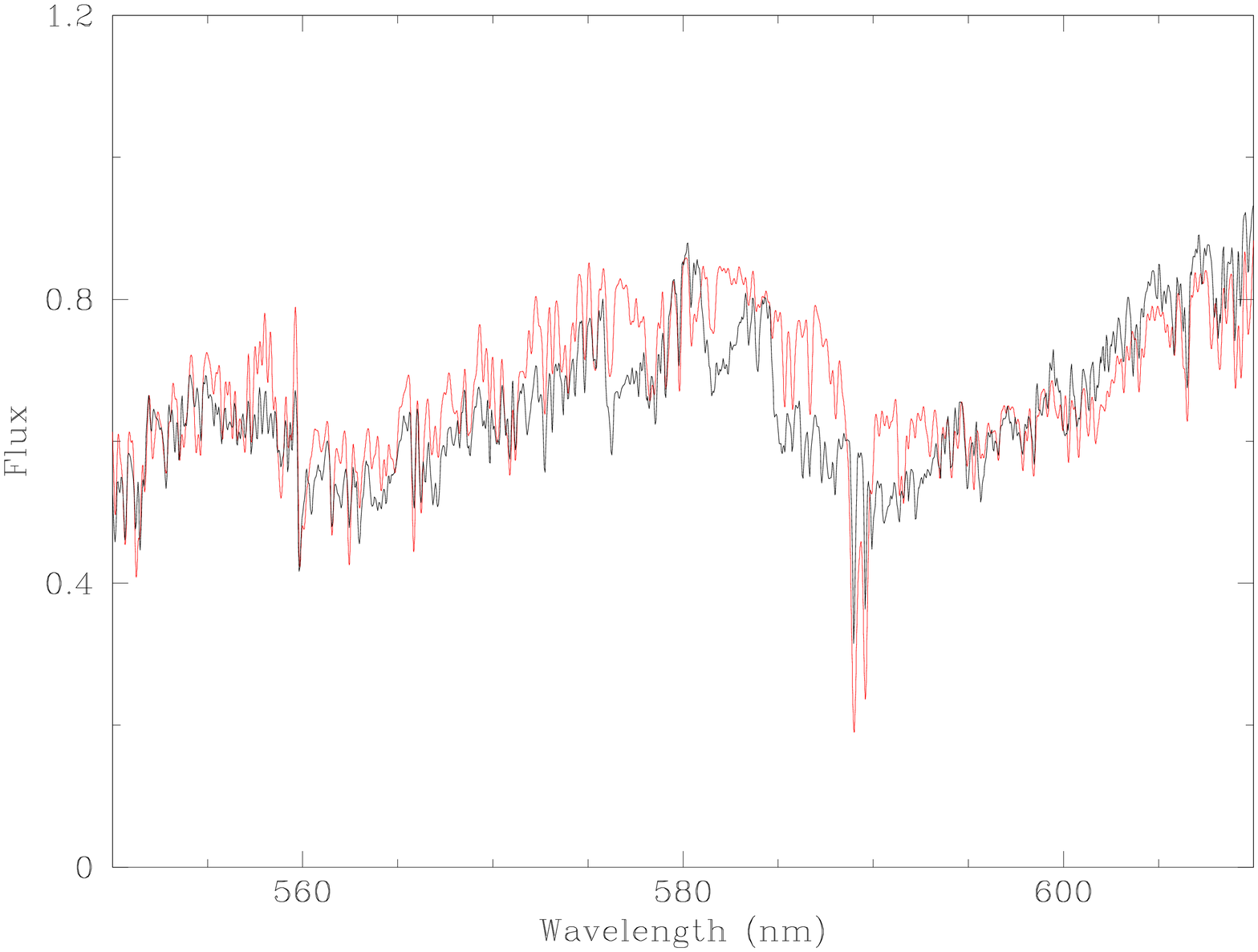} }
\caption{Comparison between the star HD 217906 (thin line), atmospheric 
parameters (T$_{\rm eff}$, log g, [Fe/H]) = (3600, 1.2, -0.11), obtained from 
Indo-US library and a synthetic spectrum (thick line), parameters (T$_{\rm 
eff}$, log g, [Fe/H], [$\alpha$/Fe]) = (3600, 1.0, 0.0, 0.0). This wavelength 
range illustrates basically the TiO $\beta$ system.}\label{fig_tio_4}
\end{figure*}

The line list for the molecules MgH, C$_2$ and CN-red 
were adopted from  Balfour \& Cartwright (1976),
Phillips \& Davis (1968) and Davis \& Philips (1963), respectively.
For the CN-blue, NH and OH-blue systems, 
the line lists by Kurucz (1993) were adopted, as implemented in
Castilho et al. (1999). For the CH blue systems, the LIFBASE program of Luque \& Crosley (1999) 
was used.
The OH vibration-rotation line list is the one implemented in Mel\'endez, Barbuy 
\& Spite (2001). 
It is based on the laboratory work of Abrams et al. (1994), 
complemented with the theoretical line list by Goldman et al. (1998). The 
molecular gf-values of the OH lines was obtained from the Einstein coefficients 
calculated by Goldman et al. (1998). 
The CN and CO infrared line lists
were implemented in Mel\'endez \& Barbuy (1999). The CO lines are from the 
theoretical work of Goorvitch (1994) and the infrared CN lines are from
S. P. Davis,
based on the analyses by M. L. P. Rao (S. P. Davis, private
communication). The FeH (A$^4\Delta$-X$^4\Delta$) line list of rotational lines
were adopted from Philips et al. (1987) as described in Schiavon et al. (1999).
The dissociation potential values 
adopted for all molecules are reported in Table 1, together with corresponding sources.

The line list for the TiO $\gamma$ system was adopted from the theoretical
work by Jorgensen (1994). The
electronic oscillator strength was calibrated to the intensity of the
laboratory line list used in Erdelyi-Mendes \& Barbuy (1989).
For the remaining TiO electronic systems, the list by Plez (1998)
was adopted. The Plez line list contains several million lines
which makes the computations extremely time-consuming so that we decided
to eliminate very weak lines in order to speed up our calculations.
Therefore, we include only the lines which satisfy simultaneously the
following set of conditions: \textit{log gf} $\geq$ -4.0, J'' $\leq$
120, and $\nu'$, $\nu''$ $\leq$ 9, where J" is the rotational quantum number
of the lower level of the transition and $\nu'$, and $\nu''$ are the vibrational
quantum numbers of the upper and lower levels of the transition. These
values were chosen after several tests were performed in order to make
sure that removal of lines not satisfying these criteria would have
a negligible impact on our computations.

\begin{table*}
\centering
\caption{Molecules considered in the computations of the synthetic spectra. References:
(1) Huber \& Herzberg (1979); (2) Pradhan et al. (1994); (3) Schulz \& Armentrout (1991);
}
\label{tab_mol}
\begin{tabular}{llccccc}
\hline Molecule & System & Number   & Wavelength        & Number of & Dissociation potencial (ref) & \\ 
                &        & of lines & coverage (nm)     & vibrational bands & D$_{\circ}$ (eV)& \\
\hline
MgH            & A$^2 \Pi-$X$^2 \Sigma$    & 1945  &  404 -  609  & 13 & 1.34 (1) \\ 
C$_2$ Swan     & A$^3 \Pi-$X$^3 \Pi$       & 11254 &  429 -  676  & 35 & 6.21 (1) \\ 
CN blue        & B$^2 \Sigma-$X$^2 \Sigma$ & 92851 &  300 -  600  & 197& 7.72 (2) \\ 
CN red         & A$^2 \Pi-$X$^2 \Sigma$    & 23828 &  404 - 2714  & 74 & " \\ 
CH AX          & A$^2 \Delta-$X$^2 \Pi$    & 10137 &  321 -  786  & 20 & 3.46 (1) \\ 
CH BX          & B$^2 \Delta-$X$^2 \Pi$    & 2016  &  361 -  682  & 10 & " \\ 
CH CX          & C$^2 \Delta-$X$^2 \Pi$    & 2829  &  270 -  424  & 12& " \\ 
CO nir         & X$^1 \Sigma^+$            & 7088  & 1578 - 5701  & 63& 11.09 (1)\\ 
NH blue        & A$^3 \Pi-$X$^3\Sigma$     & 8599  &  300 -  600  & 55& 3.47 (1) \\ 
OH blue        & A$^2 \Sigma-$X$^2\Pi$     & 6018  &  300 -  540  & 46& 4.39 (1) \\ 
OH nir         & X$^2 \Pi$                 & 2028  &  746 - 2594  & 43& "\\ 
FeH            & A$^4 \Delta-$X$^4\Delta$  & 2705  &  778 - 1634  & 9 & 1.63 (3)\\ 
TiO $\gamma$   & A$^3 \Phi$-X$^3 \Delta$   & 26007 &  622 -  878  & 23 & 6.87 (1) \\ 
TiO $\gamma$'  & B$^3 \Pi-$X$^3\Delta$     & 219367&  501 -  915  & 81 & " \\ 
TiO $\alpha$   & C$^3 \Delta-$X$^3 \Delta$ & 360725&  380 -  861  & 79& "  \\ 
TiO $\beta$    & c$^1 \Phi-$a$^1\Pi$       & 91804 &  431 -  804  & 63& "  \\ 
TiO $\delta$   & b$^1 \Pi-$a$^1\Delta$     & 189019&  622 - 1480  & 66& "  \\ 
TiO $\epsilon$ & E$^3 \Pi-$X$^3\Delta$     & 253755&  641 - 1341  & 61& "  \\ 
TiO $\phi$     & b$^1 \Pi-$d$^1\Sigma$      & 105082&  665 - 1780  & 65& "  \\ 
\hline 
\end{tabular}
\end{table*}

The electronic oscillator strengths (f${\rm _{el}}$) for the TiO
systems were obtained empirically by matching band strengths
in spectra of observed cool stars with our synthetic spectra. For this
purpose, cool giants covering the effective temperature range 3000 $\leq$
T$_{\rm eff}$ $\leq$ 3800 K were selected from  the STELIB and Indo-US
libraries. In this temperature range, TiO bands dominate the line opacity
in the visible region of the spectra of M giants. Because the model
atmospheres by Castelli \& Kurucz (2003) do not go below T$_{\rm eff}$
= 3500K, we employed MARCS model atmospheres from Plez et
al. (1992) to calibrate these f${\rm _{el}}$.

\begin{figure*}
\includegraphics[width=13cm,angle=270]{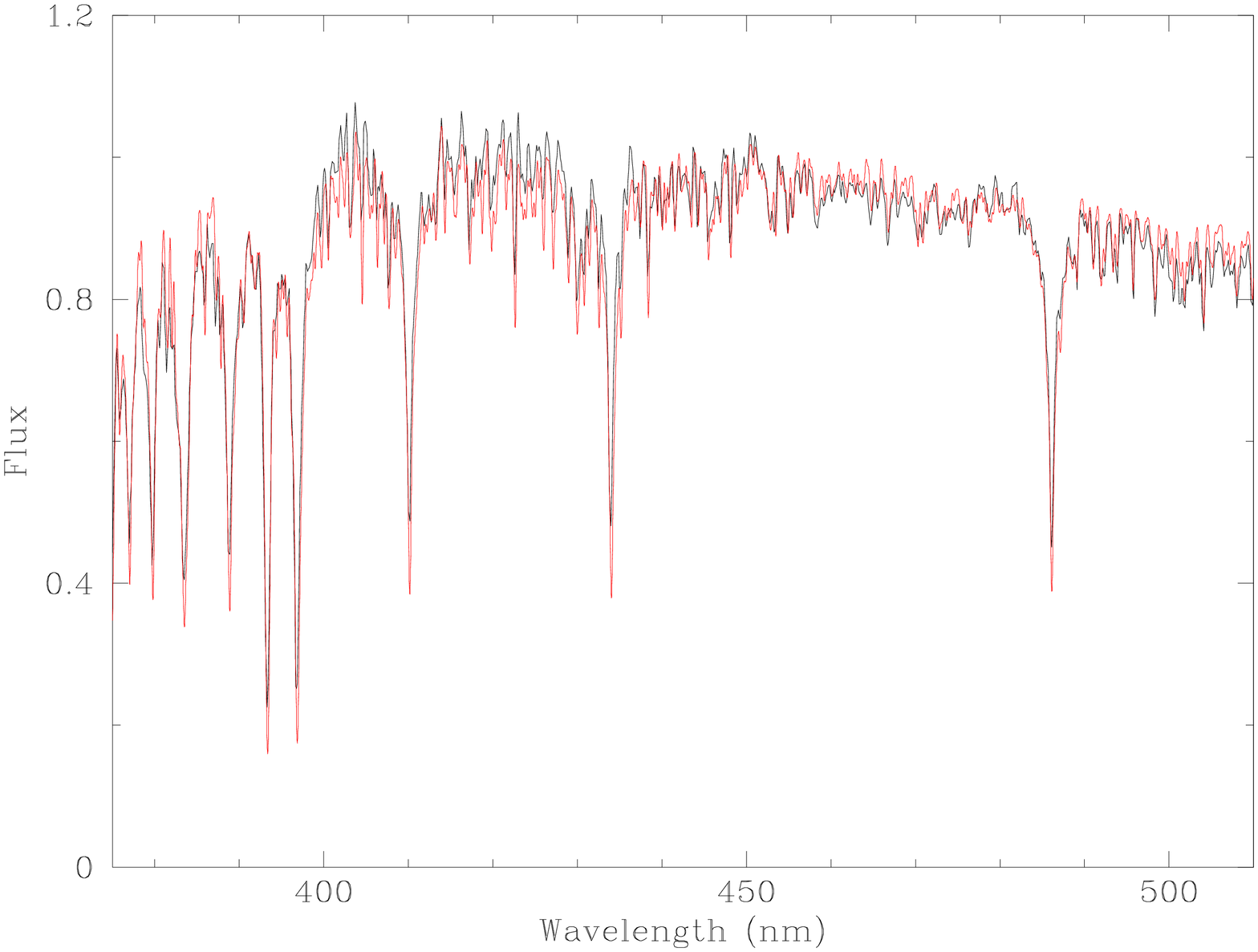} 
\caption{Comparison between the star HD 126141 (thin line), spectral type F5V 
and atmospheric parameters (T$_{\rm eff}$, log g, [Fe/H]) = (6670, 4.3, -0.02), 
obtained from STELIB library and a synthetic spectrum (thick line), parameters 
(T$_{\rm eff}$, log g, [Fe/H], [$\alpha$/Fe]) = (6750, 4.5, 0.0, 0.0). }
\label{fig_f_dwarf}
\end{figure*}

\begin{figure*}
\includegraphics[width=13cm,angle=270]{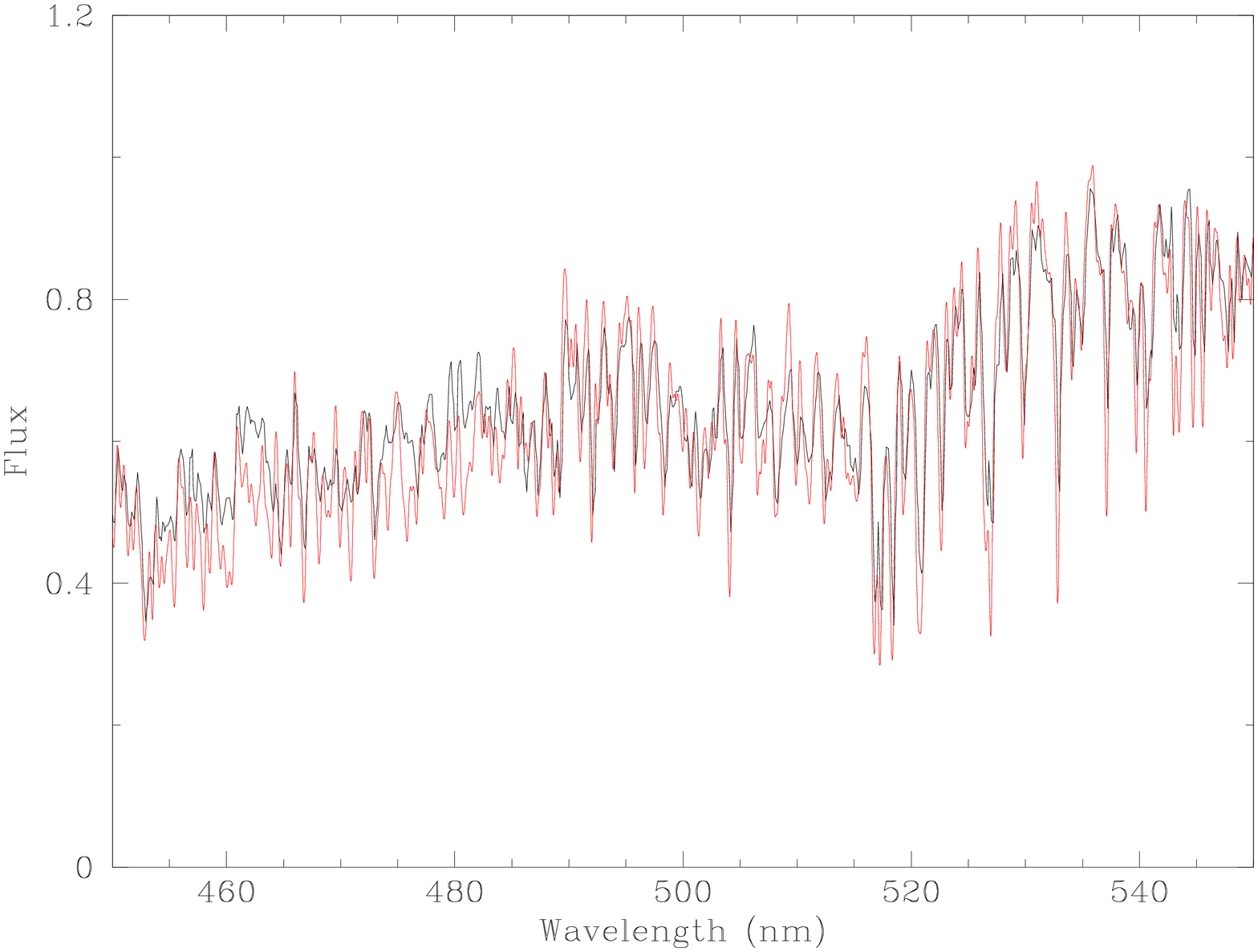} 
\caption{Comparison between the star HD 148513 (thin line), spectral type 
K4III and atmospheric parameters (T$_{\rm eff}$, log g, [Fe/H]) = (3979, 1.03, 
-0.14), obtained from STELIB library and a synthetic spectrum (thick line), 
parameters (T$_{\rm eff}$, log g, [Fe/H], [$\alpha$/Fe]) = (4000, 1.0, 0.0, 
0.0). }\label{fig_k_giant}
\end{figure*}

In Figure \ref{fig_tio0} the relative contribution of each of the TiO
systems, calculated for a cool giant, is presented. 
Some comparisons between empirically calibrated TiO-dominated synthetic spectra
to the spectra of some observed stars are presented in Figures
\ref{fig_tio_1} to \ref{fig_tio_4}. 
Moreover, the comparison to the spectra of an F dwarf 
and a K giant are presented in Figures \ref{fig_f_dwarf} and \ref{fig_k_giant} respectively. The
wavelength range plotted in the later two figures were chosen to illustrate the spectral region where
those spectral types dominate the flux of integrated population spectra.
The synthetic spectra shown in these figures
are the ones whose atmospheric parameters are the closest
to those of the observed stars, as given 
in the respective library (STELIB or Indo-US).

In all figures from \ref{fig_tio_1} to \ref{fig_k_giant} the
synthetic spectra were convolved to match the different resolutions
of the observed spectra (FWHM $\approx$ 1 \AA\ for Indo-US and
$\approx$ 3 \AA\ for STELIB).

\section{Library of synthetic spectra}

Given the atomic and molecular line lists described in the previous
section, a library of synthetic spectra was computed from 3000 \AA\
to 1.8 ${\rm \mu}$m, in steps of 0.02${\rm\AA}$. 
The model atmosphere grid is available for
microturbulent velocities v$_{\rm t}$ = 2 km/s, but the synthetic spectra
were calculated with values more representatives of observed stars
as given below:

\begin{itemize}
\item v$_{\rm t}$ = 1.0 km s$^{-1}$ for log $g$ $\geq$ 3.0;
\item v$_{\rm t}$ = 1.8 km s$^{-1}$ for 1.5 $\leq$ log $g$ $\leq$ 2.5, and;
\item v$_{\rm t}$ = 2.5 km s$^{-1}$ for log $g$ $\leq$ 1.0.
\end{itemize}

The output of the synthesis code is normalized flux. In order to
produce spectra in an absolute flux scale, the normalized
high-resolution spectra were multiplied by the true continuum given
directly by the ATLAS9 model atmosphere.  The library is presented in both
normalized and absolute flux formats, and each spectrum is stored
in a FITS file consisting of two apertures ({\it multispec} IRAF format),
where aperture number 1 is the normalized spectrum and aperture number 2
is an absolute flux spectrum. The spectral library is available
upon request.

\subsection{Coverage in Stellar Parameter Space}

The main grid covers the  following parameters:

\begin{itemize}
\item Effective temperatures: 3500 $\leq $ T$_{\rm eff}$ $\leq $ 7000 K in steps 
of 250K\item Surface gravities: 0.0 $\leq $ log g $\leq $ 5.0 in steps of 0.5
\item Metallicities: ${\rm [Fe/H]}$ = -2.5, -2.0, -1.5, -1.0, -0.5, 0.0, 0.2 and 
0.5\item Chemical compositions: ${\rm [\alpha/Fe]}$ = 0.0 and 0.4, where the 
$\alpha$-elements considered are O, Ne, Mg, Si, S, Ca and Ti.\end{itemize}

It is not well established yet whether titanium is 
associated to the synthesis of the $\alpha$-elements or to the iron-peak 
elements (Timmes et al. 1995). Nevertheless, it has been
shown that its abundance follows the same pattern as that of
$\alpha$-elements for several types of populations (e.g. McWilliam \& Rich 1994; 
Pomp\'eia et al. 2003), and thus
we decided to include this element in the $\alpha$-enhanced group.

Concerning calcium, observations of halo and disk Galactic stars show that [Ca/Fe]
follows closely the behaviour of other $\alpha$-elements as a
function of [Fe/H] (McWilliam \& Rich 1994; Edvardsson et al. 1993).
In the bulge of our galaxy, however, some high-resolution analyses show that 
[Ca/Fe] seems to be close to
solar (e.g. Zoccali et al. 2004). From lower resolution 
studies, there is evidence as well that [Ca/Fe] is not enhanced 
in elliptical galaxies (e.g. Thomas et al. 2003b), although Prochaska, Rose \&
Schiavon (2005) showed that the calcium abundance derived from the Ca4227 index is dependent
on the index-measuring strategy.

\begin{figure}
\includegraphics[width=8.5cm]{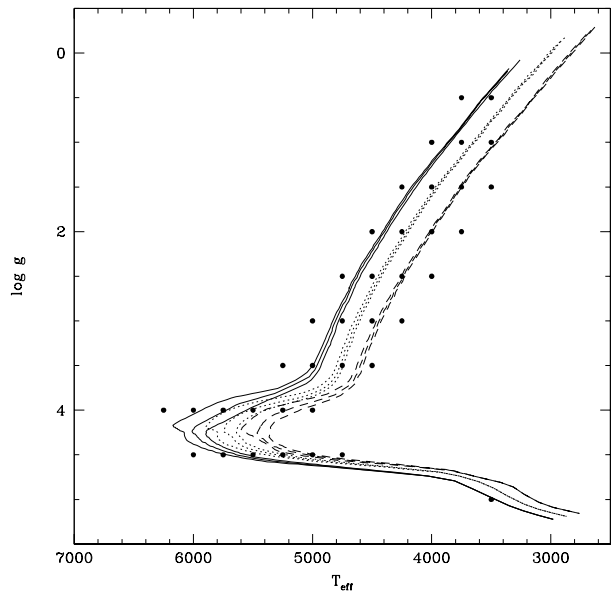} 
\caption{Bertelli et al. (1994) isochrones for three metallicities ([Fe/H] = -0.4, 0.0, 
+0.4; solid, dotted and dashed lines) and three ages (8, 10.5 and 13 Gyr; from 
left to right). The circles indicate the extension of the $\alpha$-enhanced main 
grid, computed with [$\alpha$/Fe] = 0.4, [Ca/Fe] = 0.0 and [Fe/H] = (-0.5, 0.0, 0.2 and 
0.5). }
\label{fig_padova}
\end{figure}

Considering the application of this grid to the study of old and
metal-rich stellar populations, a subsample of the $\alpha$-enhanced
main grid was also computed with [Ca/Fe] = 0.0, for [Fe/H] = -0.5,
0.0, 0.2 and 0.5. The sets of T$_{\rm eff}$ and log g for these latter
computations were chosen amongst those that are relevant for old stellar
populations at these metallicities, for ages between 8 and 13 Gyr. 
Bertelli et al. (1994) isochrones were used to guide our selection. 
Some of these isochrones are presented
in Figure \ref{fig_padova} as well as the (T$_{\rm eff}$, log g) pairs
computed for this extension of the library.

\begin{figure*}
\includegraphics[width=16cm]{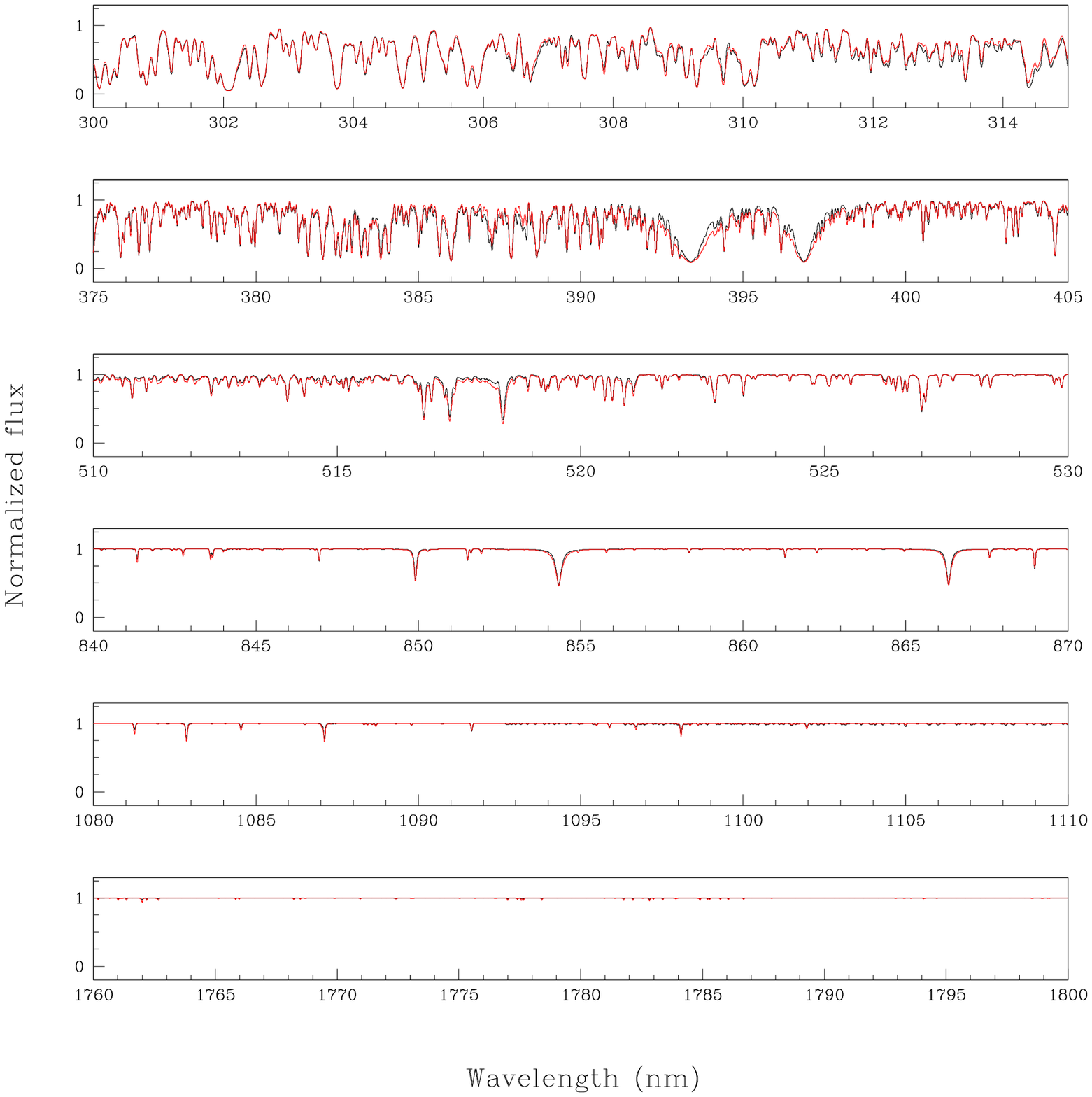} 
\caption{Solar scaled (thin line) and $\alpha$-enhanced (thick line) synthetic spectrum computed with
T$_{\rm eff}$ = 4500K, log g = 3.0 and [Fe/H] = -2.0, for different wavelength regions.
The spectra were convolved to FWHM $\sim$ 0.5 ${\rm \AA}$.}
\label{fig_alfa1}
\end{figure*}

\begin{figure*}
\includegraphics[width=16cm]{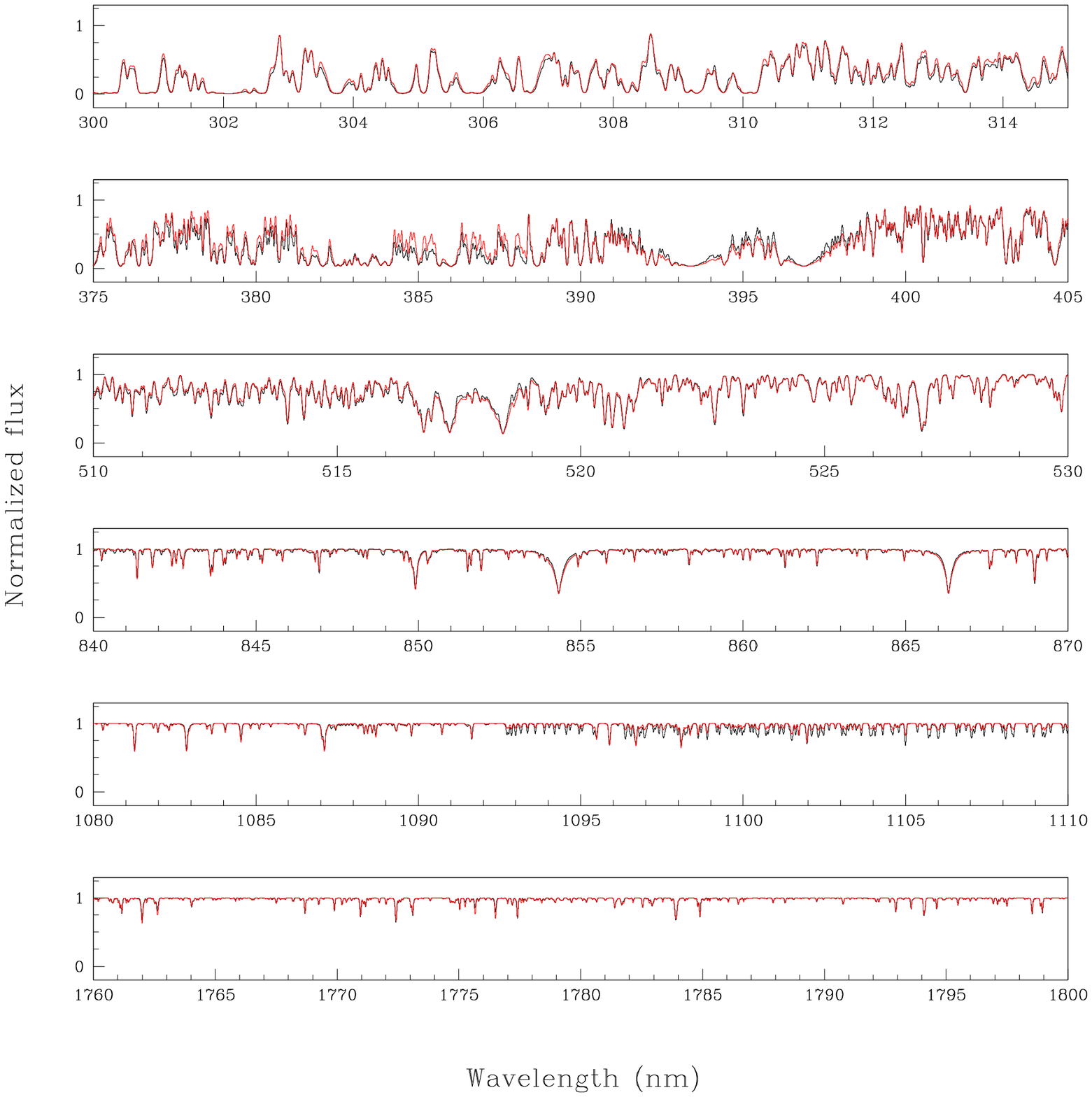} 
\caption{Solar scaled (thin line) and $\alpha$-enhanced (thick line) synthetic spectrum computed with
T$_{\rm eff}$ = 4500K, log g = 3.0 and [Fe/H] = 0.0, for different wavelength regions.
The spectra were convolved to FWHM $\sim$ 0.5 ${\rm \AA}$.}
\label{fig_alfa2}
\end{figure*}

High resolution synthetic spectra computed for a cool giant with
solar-scaled and $\alpha$-enhanced abundance patterns are compared in
\ref{fig_alfa1} and \ref{fig_alfa2}, for the metal-poor and metal-rich cases, respectively.
As expected, well known indicators of abundances of $\alpha$-elements, such
as the Mg triplet lines at 517 nm, the MgH band at 518 nm and the Ca
triplet in the far red, are stronger in the $\alpha$-enhanced spectra. We
note however other lines which are very sensitive to $\alpha$-enhancement and
yet are virtually unexplored in stellar populations work (e.g. the Ca H \& K
lines and the lines due
to the CN molecule, at $\lambda$ $>$ 1093 nm). 
We also note that some lines
are visibly weaker in the $\alpha$-enhanced case (e.g. the region around 385 nm in
Figure \ref{fig_alfa2}).
This cannot be attributed to individual abundance variations of the non-$\alpha$ elements,
because those abundances are the same as in the solar scaled case. 
The decrease of the intensity of some lines can be explained because the $\alpha$-elements, 
specially Mg, are electron donors 
and thus contribute in an important way to the continuum opacity. 
Because the spectra shown in Figures \ref{fig_alfa1} and \ref{fig_alfa2}
are normalized, the net
effect is that some lines in the $\alpha$-enhanced case look weaker due to
enhanced continuum opacity, even when the individual abundances corresponding 
to those lines are unchanged.

\subsection{The relative contribution of different chemical species to
the total line opacity}

A great advantange of synthetic spectra over
observed spectra, is that with spectral synthesis
it is possible to disentangle the relative
contribution of different chemical species to
the line spectrum, which allows the study of line
indices in a more careful way. For example, it is
possible to evaluate the presence of blends or the 
range of applicability of
a given line indice.

\begin{figure*}
\includegraphics[width=16cm]{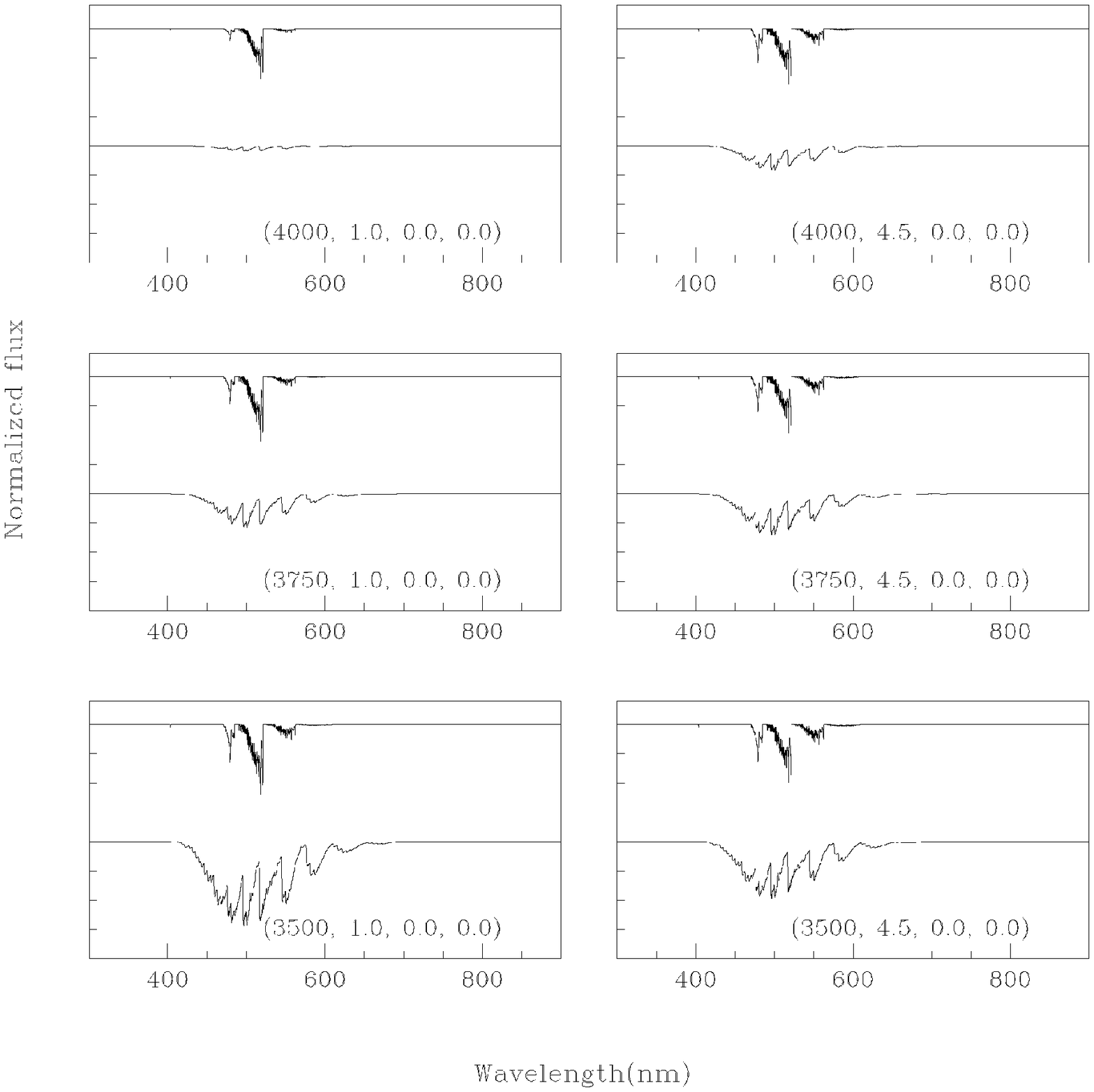} 
\caption{
Individual contribution of the MgH and TiO $\alpha$ molecules, for three T$_{\rm 
eff}$ and two log g (indicated in the figures). The top plot in each panel 
represents the MgH spectrum, and the bottom plot presents the TiO $\alpha$ 
spectrum.}\label{fig_mc6}\end{figure*}

As an example, the well-known Lick indices Mg$_1$, Mg$_2$ and Mg$_b$
 measure the
intensity of the MgI triplet lines (5167.327, 5172.698 and
 5183.619 ${\rm \AA}$) together with MgH, C$_2$ and TiO bands,
besides other atomic lines. These indices measure essentially
MgI and MgH in K giants and in the integrated light of galaxies.
However, it should be noted that in stars cooler than 3750 K,
bands from the $\alpha$ system of TiO become stronger than MgH, as can be seen in
 Figure \ref{fig_mc6}. 
 It is clear that for very metal-rich stars at the tip of the red giant
branch and galaxies containing such stars these indices are substantially
affected by TiO bands, which makes their interpretation somewhat more complicated 
(see also Schiavon 2005).
 
\begin{figure*}
\includegraphics[width=16.5cm]{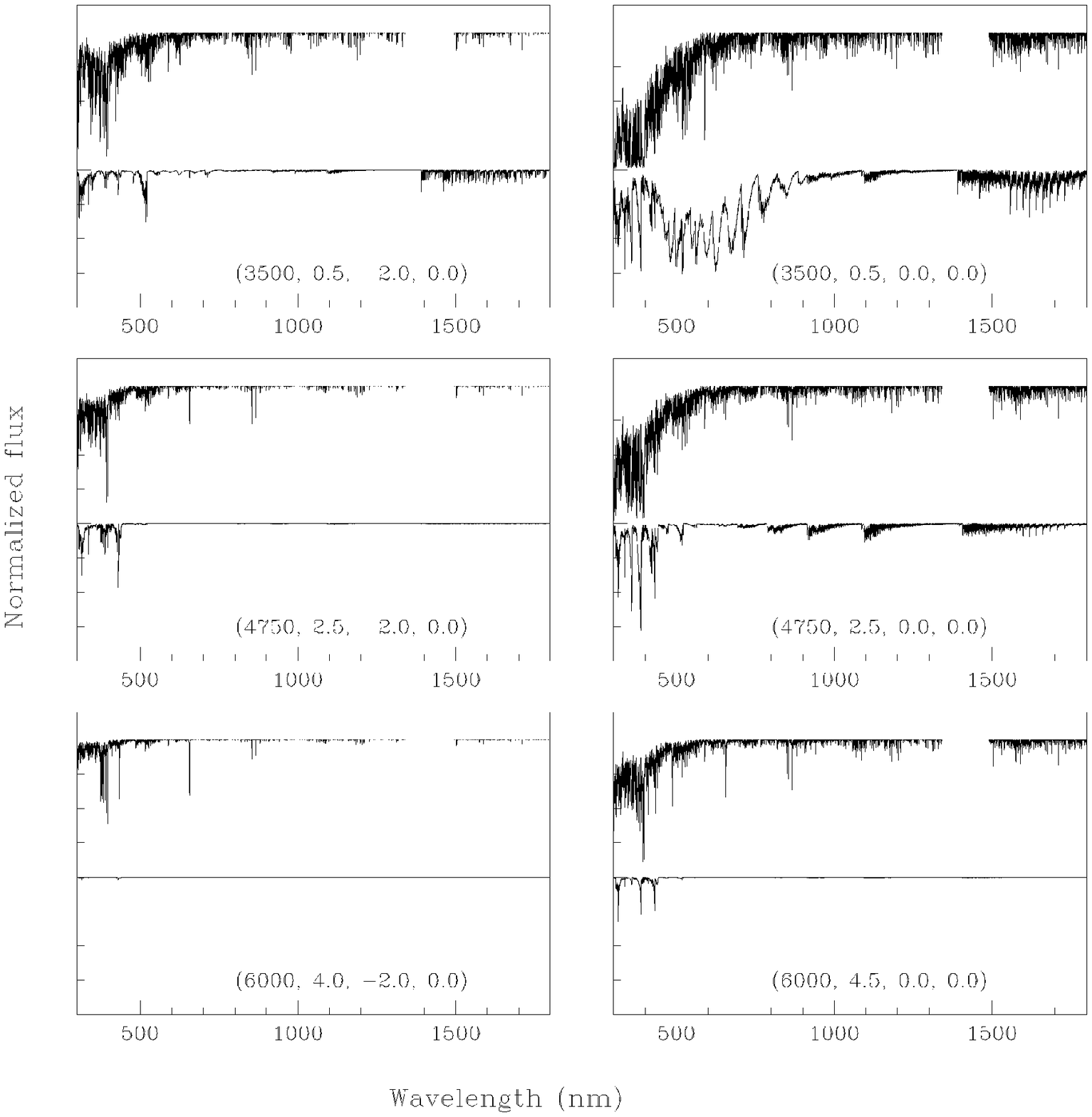} 
\caption{The relative contribution of atomic and molecular line opacities for 
six combinations of (T$_{\rm eff}$, log g, [Fe/H], [$\alpha$/Fe]) as indicated
in each panel. 
The top spectrum of each panel shows the normalized atomic contribution (an offset
was applied) and the lower 
spectrum shows the normalized molecular contribution.}
\label{fig_mc1}
\end{figure*}

\begin{figure*}
\includegraphics[width=16.5cm]{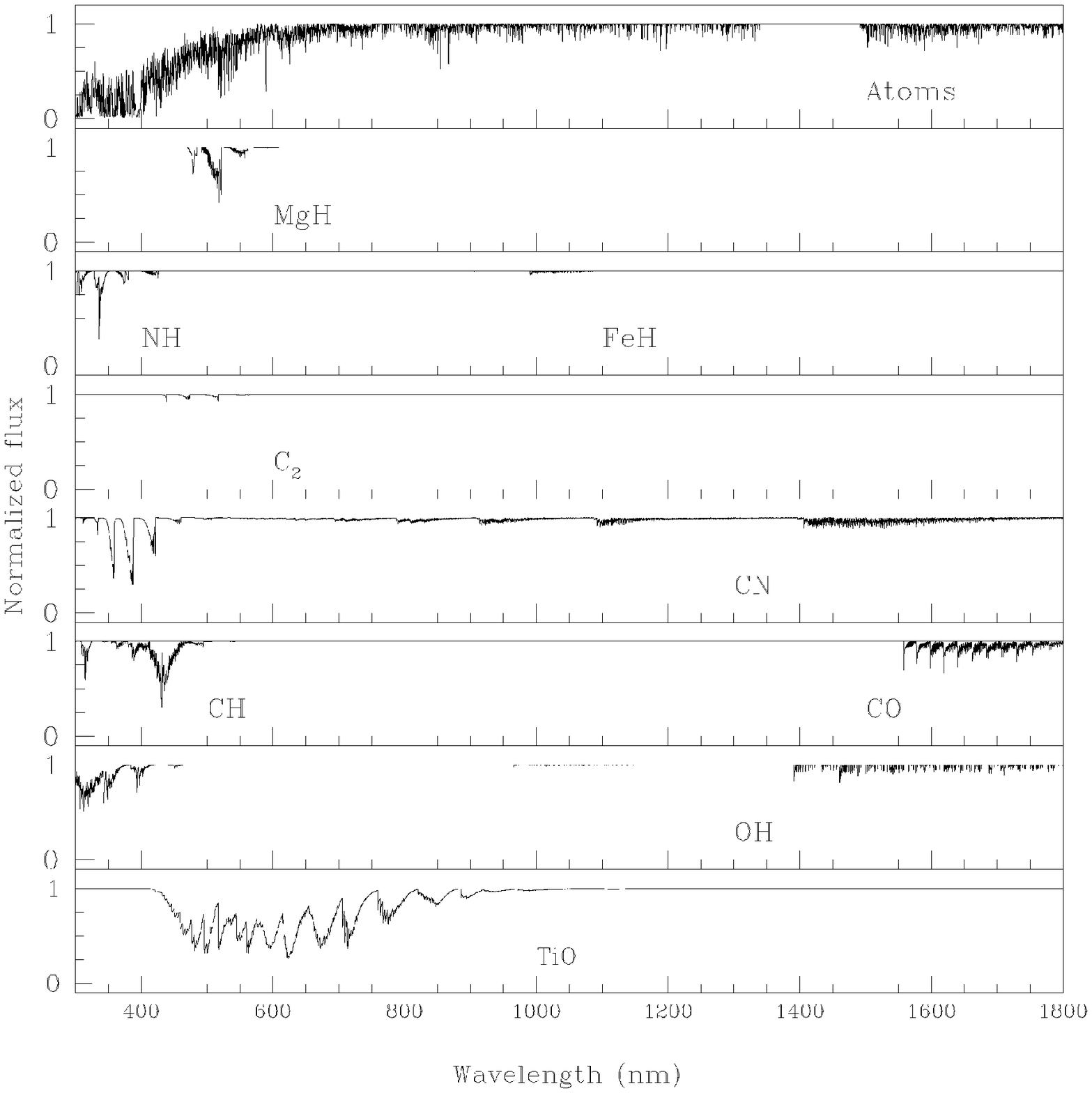} 
\caption{
Individual contribution from each molecule for the parameters (T$_{\rm eff}$, 
log g, [Fe/H], [$\alpha$/Fe]) = (3500, 0.5, 0.0, 0.0). The molecules considered 
are labeled in each panel. }\label{fig_mc3}
\end{figure*}

\begin{figure*}
\includegraphics[width=16.5cm]{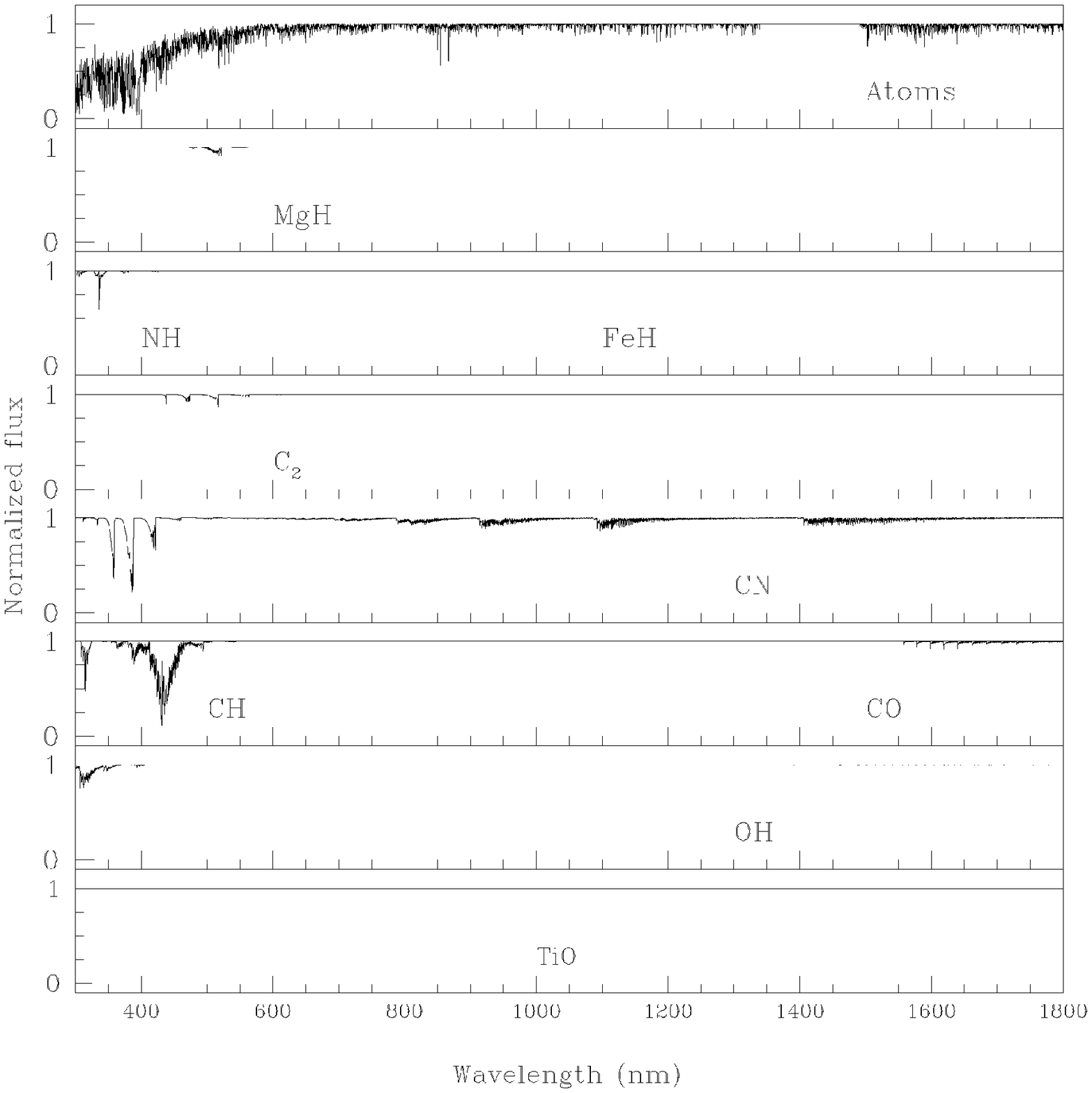} 
\caption{
Same as Figure \ref{fig_mc3} for (T$_{\rm eff}$, log g, [Fe/H], [$\alpha$/Fe]) = 
(4750, 2.5, 0.0, 0.0).}
\label{fig_mc4}
\end{figure*}

\begin{figure*}
\includegraphics[width=16.5cm]{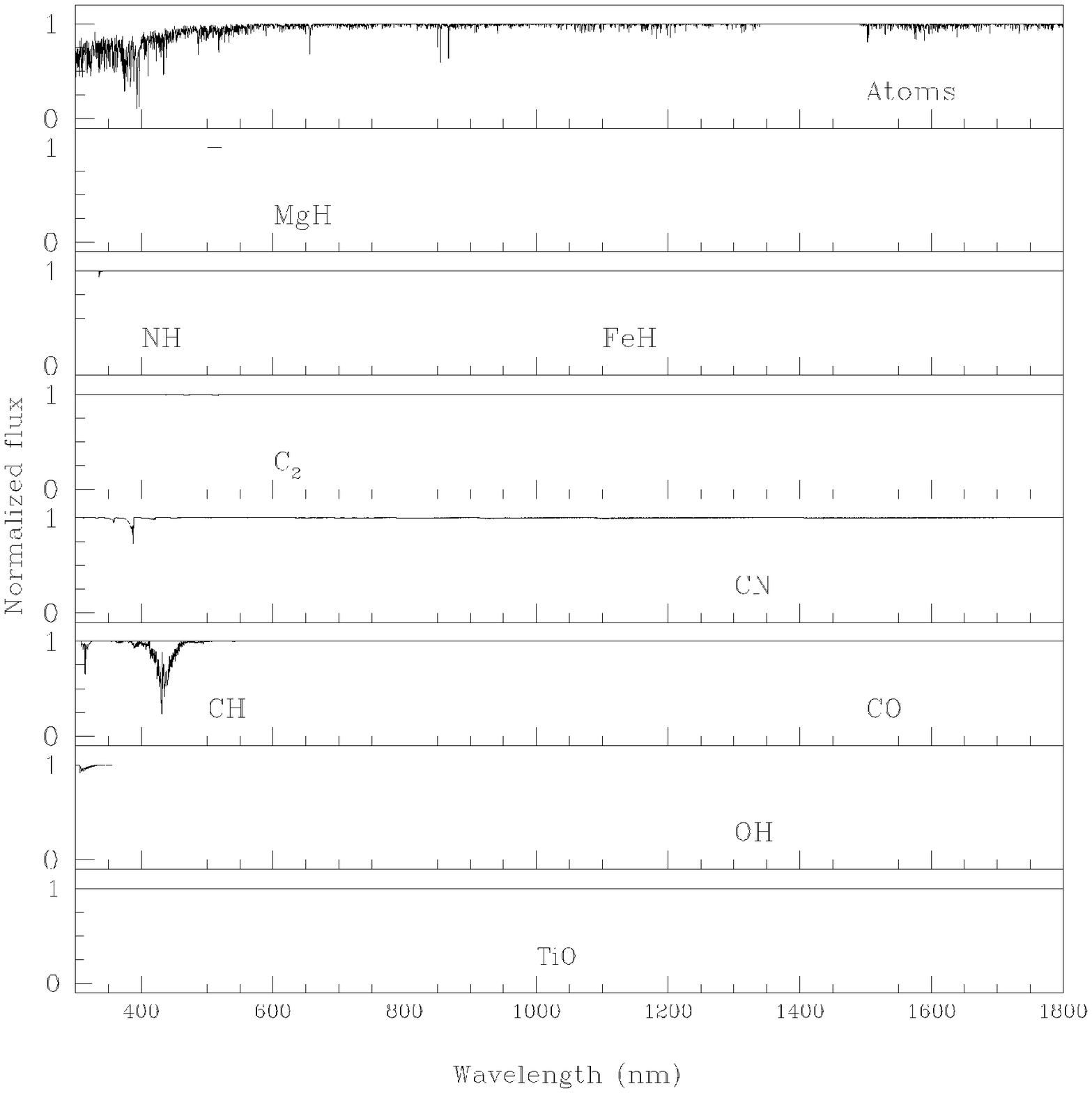} 
\caption{
Same as Figure \ref{fig_mc3} for (T$_{\rm eff}$, log g, [Fe/H], [$\alpha$/Fe]) = 
(6000, 4.0, 0.0, 0.0). Only the molecules CH and CN have non-negligible effect 
in the total spectrum.}\label{fig_mc5}
\end{figure*}

The separate contribution of atoms and molecules
is shown in Figure \ref{fig_mc1} for the parameters (T$_{\rm eff}$, log g)  = (3500, 0.5),
(4750, 2.5) and (6000, 4.0) for [Fe/H] = -2.0 and 0.0 ([$\alpha$/Fe]
= 0.0). 
In order to illustrate the contribution of atoms and each
diatomic molecule, we computed each molecule separately.
The results are shown in Figures \ref{fig_mc3} - \ref{fig_mc5}, where the
impact of each molecule on the line spectrum can be
evaluated.

\section{Summary}

A library of synthetic stellar spectra is presented. This library
was computed with a sampling 0.02\AA\ and for the parameters: effective
temperatures 3500 $\leq$ T$_{\rm eff}$ $\leq$ 7000 K,  surface gravities
0.0 $\leq$ log g $\leq$ 5.0, metallicities -2.5 $\leq$ [Fe/H] $\leq$ +0.5,
and $\alpha$-elements over Fe ratio [$\alpha$/Fe] = 0.0 and 0.4. For the
metallicites [Fe/H] $\geq$ -0.5, two $\alpha$-enhanced grids are available,
with and without Ca in the enhanced group.

For the computation of this grid, an atomic line list was implemented
based on the previous lists  of Barbuy et al. (2003) and Mel\'endez \& Barbuy
(1999). The \textit{log gf} values were updated to the most recent NIST values,
and damping constants were calculated for the totality of strong lines
following ABO papers.

The molecular line opacities include the molecules MgH, C$_2$, CN, CH, CO, OH,
NH, FeH and TiO. The TiO systems, which dominate the opacity of cool
stars in the visible wavelength range, were calibrated by comparison
with observed stars in the empirical libraries STELIB and Indo-US.

This high-resolution stellar synthetic library covers simultaneosly a
wide wavelength range, and $\alpha$-enhanced chemical
compositions. The $\alpha$-enhancement in particular is indispensable to
the reproduction of the integrated properties of old stellar populations,
for all metallicities regimes. This library is a valuable and powerfull
tool for the study of stellar populations, and can be used to model old
metal-rich stellar populations with unprecedented accuracy.

{\it Note added after acceptance}:
We would like to mention that by the time of the acceptance of our paper, 
two recent libraries including variable $\alpha$-enhancement came to our knowledge, Munari 
et al. (2005) and Brott \& Hauschild (2005).

\begin{acknowledgements}
We are grateful to the referee Robert Kurucz for his comments,  
Roger Cayrel for valuable discussions
on collisional broadening, and Dinah M. Allen
for the heavy-elements line list.
PC acknowledges: a Fapesp PhD fellowship n$^{\circ}$ 2000/05237-9;
the Latin American-European Network on Astrophysics and Cosmology
(LENAC) of the European Union's ALFA Programme; and 
Fiorella Castelli, for the help with the ATLAS model atmospheres.
BB  acknowledges partial 
financial support from CNPq and Fapesp. R.P.S. acknowledges financial support
from HST Treasury Program grant GO-09455.05-A to the University of Virginia.
\end{acknowledgements}

\appendix
\section{Calculation of collisional broadening}
 
\begin{figure}
\includegraphics[width=8.8cm]{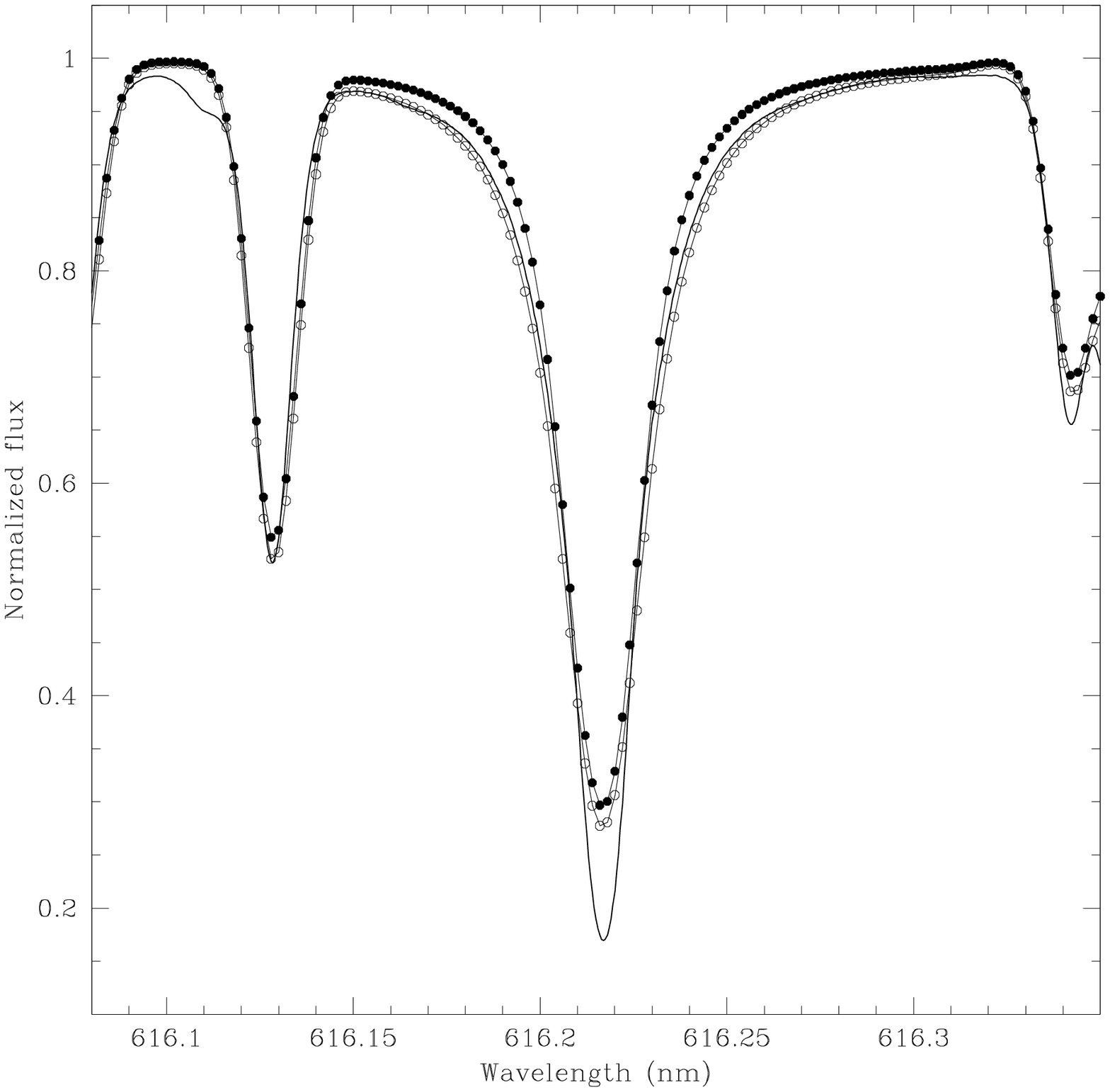} 
\caption{Comparison between the solar spectrum (solid line, no symbols) and synthetic 
spectrum calculated with ATLAS9 (solid line, open circles) and HM74 (solid line, filled symbols) model 
atmospheres, around the Ca 6162 \AA\ line. The same parameters were adopted for 
the atomic transition considered in both calculations, and thus the difference 
in the line profiles is due only to the different model atmospheres 
employed.}\label{fig_ca}
\end{figure}

The main source of
collisional broadening for metallic lines in stars of spectral
types between F and M is  collisions with neutral hydrogen. The
spectral synthesis code evaluates $\gamma$ for each atmospheric
layer given the interaction constant $C_6$:

\begin{equation}
\gamma_6/N_H = 17 v^{3/5} C_6^{2/5}
\label{eq_gamma} 
\end{equation}

where $v$ is the relative velocity between the colliding particles 
and $N_H$ is the density of Hydrogen atoms.
The $C_6$ parameter (or any other parameter related to
the damping width) is accurately known only for very few lines. The
Unsold (1955, see also Gray 1976) approximation allows one to straightforwardly
estimate the damping constants, but it has long been recognized that
it underestimates the collisional broadening, and thus empirical
enhancement factors are applied to $C_6$ in order to reproduce
the observed line profiles.

The works by O`Mara (1976) and Anstee \& O'Mara (1991) developed an alternative
theory for the computation of cross sections $\sigma$ of
atomic transitions, thus providing a more accurate way to evaluate $
\gamma$. Anstee \& O'Mara (1995), Barklem \& O'Mara (1997) and
Barklem, O'Mara \& Ross (1998) published cross sections for a wide
range of \textit {s-p} and \textit {p-s}, \textit {p-d} and
\textit {d-p},
\textit {d-f} and \textit {f-d} transitions, respectively. Herefrom,
we refer to these cross-sections as ABO.

In order to use the more precise ABO theory in our calculations,
we had to relate the ABO cross section $\sigma $ to the classical
interaction constant $ C_6 $ which is the input value to our
spectrum synthesis code.  The interaction constant can be expressed by:

\begin{equation}
C_6=6.46\times10^{-34} (R_{hi}^{2} - R_{lo}^{2})
\label{eq_c6}
\end{equation}

where $R_{hi}^{2}$ and $R_{lo}^{2}$ are the mean-square radii
of the upper and lower states of the transition in atomic units. The
cross section from the van der Waals theory is given by:

\begin{equation}
\sigma_{vdW} = 63.65 (R_{hi}^{2} - R_{lo}^{2})^{2/5}
\label{eq_sigma}
\end{equation}

From equations \ref{eq_c6} and \ref{eq_sigma}, and replacing the classical
cross-section $\sigma _{vdW}$ by the cross-section given by ABO theory 
$\sigma_{ABO}$, we obtain the relation:

\begin{equation}
C_6=6.46\times 10^{-34}(\sigma _{ABO}/63.65)^{5/2} 
\label{eq_c6_abo}
\end{equation}

Or in terms of $\gamma_6$:

\begin{equation}
\gamma_6/N_H=1.415\times 10^{-14}v^{3/5}\sigma _{ABO}
\label{eq_g6_abo}
\end{equation}

The expression in Eq. \ref{eq_gamma} leads to a $\gamma \propto T^{2/5}$
relationship. However, according to the theory developed by ABO, which is based on
the formalism by Brueckner (1971), $\gamma$ has a dependence on the temperature
that varies between $
T^{0.3} $ and $ T^{0.4} $. The ABO papers tabulate the velocity exponents
$\alpha$ that take into account this non-universal dependence. Due to
the different formalisms used by ABO theory and by the spectral synthesis
code, the dependences of $\gamma_6$ on temperature are
somewhat different. Nevertheless, the error introduced
in the computed line profile is negligible ($<$ 5\%) given
other sources
of errors like uncertainties in the 
atmospheric parameters, model atmospheres and NLTE effects.

The $ \sigma_{ABO} $ values were obtained primarily from Barklem et
al. (2000). Cross sections for lines not included in Barklem et al. (2000)
were calculated through the code presented in Barklem, O'Mara \& Ross (1998).
For this purpose,
the energy levels and orbital angular momentum quantum
numbers for each transition were obtained from the
line list of Kurucz (1993), available at the web address
http://kurucz.harvard.edu/linelists.
The ionization limits of the transitions were approximated by the
first ionization potential of the atom. Although this approximation is not
strictly accurate, as pointed out in Barklem et al. (2000), the detailed
inclusion of the ionization limits of each transition is beyond the
scope of the present work.

A further step in critically evaluating the damping widths of the
strongest lines is to take into account the model atmosphere adopted.
This effect is
illustrated for the Ca line 6162\AA\ in Figure \ref{fig_ca}. Cross
sections for three lines of Ca I (${\rm \AA}$ = 6102, 6122, 6162)
have been accurately calculated by Spielfiedel et al (1991). The ratios
$\gamma_{ABO}/\gamma_{Spielfiedel}$ for these lines are (0.8, 1.05,
0.96), which indicates good agreement.
In Figure \ref{fig_ca}
is shown the solar spectrum (solid line), a synthetic spectrum
calculated using the solar model atmosphere by Holweger \& Muller (1974,
HM74, filled circles), and another synthetic spectrum calculated
with ATLAS9 models (open circles). The $C_6$ employed is the one
directly obtained from equation \ref{eq_c6_abo}, using the same
parameters as given in Anstee \& O`Mara (1995), Table 3.

It can be seen that the line profile computed with the
ATLAS9 model is stronger than when the HM74 model is used.
The use of a MARCS model atmosphere (Plez et al.
1992) results in a line profile that is very similar to that
obtained when the ATLAS9 model is employed.

The computed line profile depends on both the damping constant
and the model atmosphere. Hence, even employing perfect
damping parameters does not guarantee that correct
profiles are computed, since the synthetic profile
also depends on the adopted model atmosphere, as already
pointed out by Cayrel et al. (1996) and Barbuy et al. (2003),
and as is clearly shown in Fig. A1.

As a result, 5\% of the lines for which the broadening values were calculated
through ABO theory have too strong wings when compared with the observed
spectrum of the Sun, when calculated with an ATLAS9 model.
Therefore, the $ C_6 $ values for these lines
were fitted manually to the solar spectrum.  It was found that the average
$\gamma$(ABO)/$\gamma$(best fit) for those lines is $\approx$ 1.4. Cayrel \&
Van't Veer (private communication) performed similar tests for the
Mg I triplet lines, and found that $\gamma$(ABO)/$\gamma$(best fit)
$\approx$ 1.5, which is in remarkable agreement with our own results.

The wings of the weaker lines 
are not critically affected by the use of different model
atmospheres (as can be seen for the two weaker lines in Figure
\ref{fig_ca}). This implies that for $\approx$ 95\% of the lines,
the broadening predicted by ABO theory fits well the solar spectrum when
ATLAS9 models are employed. 



\begin{thebibliography}{}


\bibitem[]{} Abrams, M. C., Davis, S. P., Rao, M. L. P., Engleman, R., Jr., 1994, ApJS, 93, 351
\bibitem[]{} Allard, F. \& Hauschildt, P.H. 1995, ApJ, 445, 433
\bibitem[]{} Allen, D. M., 2005, PhD Thesis, Univ. de S\~ao Paulo
\bibitem[]{} Anstee, S.D., O'Mara, B.J., 1991, MNRAS, 253, 549
\bibitem[]{} Anstee, S.D., O'Mara, B.J., 1995, MNRAS, 276, 859

\bibitem[]{} Barbuy, B. 1982, PhD Thesis, Universit\'e de Paris VII
\bibitem[]{} Barbuy, B., Perrin, M.-N., Katz, D., Coelho, P., Cayrel, R., Spite, M., Van't Veer-Menneret, C., 2003, A\&A, 404, 661
\bibitem[]{} Balfour, W. J., Cartwright, H. M., 1976, A\&AS, 26, 389
\bibitem[]{} Barklem, P.S., O'Mara, B.J. 1997, MNRAS, 290, 102
\bibitem[]{} Barklem, P.S., O'Mara, B. J., 1998, MNRAS, 300, 863
\bibitem[]{} Barklem, P.S., O'Mara, B.J., Ross, J.E. 1998, MNRAS, 296, 1057
\bibitem[]{} Barklem, P. S., Piskunov, N., O'Mara, B. J., 2000, A\&AS, 142, 467
\bibitem[]{} Bell, R. A., Dwivedi, P. H., Branch, D., Huffaker, J. N., 1979, ApJS, 41, 593
\bibitem[]{} Bensby, T., Feltzing, S., Lundstrom,öI., 2003,A\&A, 410, 527
\bibitem[]{} Bertelli, G., Bressan, A., Chiosi, C., Fagotto, F., Nasi, E., 1994, A\&AS, 106, 275
\bibitem[]{} Bessell, M. S., 1991, A\&AS, 89, 335.
\bibitem[]{} Bessell, M. S., Brett, J. M., Wood, P. R., Scholz, M., 1989, A\&AS, 77, 1
\bibitem[]{} Biehl, D., 1976, PhD thesis, Univ. Kiel
\bibitem[]{} Brott, I., Hauschildt, P.H., 2005, in The Three Dimensional Universe with Gaia, M.A.C. Perryman and C. Turon eds., ESA SP-576, 565
\bibitem[]{} Bruecknerm K., 1971, ApJ, 169, 621
\bibitem[]{} Bruzual, A. G., Charlot, S., 1993, ApJ, 405, 538
\bibitem[]{} Bruzual, A. G., Charlot, S., 2003, MNRAS, 344, 1000
\bibitem[]{} Burstein, D., Faber, S.M., Gaskell, C.M., Krumm, N., 1984, ApJ, 287, 56
\bibitem[]{} Buzzoni, A., 2002, AJ, 123, 1188


\bibitem[]{} Castelli, F., Kurucz, R. L., 2003, Proceedings of the 210th Symposium of the International Astronomical Union held at Uppsala University, Uppsala, Sweden, 17-21 June, 2002. Edited by N. Piskunov, W.W. Weiss, and D.F. Gray. Published on behalf of the IAU by the Astronomical Society of the Pacific, 2003., p.A20
\bibitem[]{} Castilho, B.V., Spite, F., Barbuy, B., Spite, M., de Medeiros, J.R., Gregorio Hetem, J. 1999, A\&A, 345, 249
\bibitem[]{} Cayrel, R., Perrin, M.-N., Barbuy, B., Buser, R., 1991b, A\&A, 247, 108
\bibitem[]{} Cayrel, R., Perrin, M. N., Buser, R., Barbuy, B., Coupry, M. F., 1991a, A\&A, 247, 122
\bibitem[]{} Cayrel, R., Faurobert-Scholl, M., Feautrier, N., Spielfieldel, A., Thevenin, F., 1996, A\&A, 312, 549
\bibitem[]{} Cervi\~no, M., Mas-Hesse, J., 1994, A\&A, 284, 749
\bibitem[]{} Chavez, K., Malagnini, M. L., Morossi, C. 1997, A\&AS, 126, 267


\bibitem[]{} Davis, S. P., Phillips, J. G., 1963, {\it Berkeley Analyses of Molecular Spectra}, Berkeley: University of California Press
\bibitem[]{} Delgado, R. M. Gonz\'alez, Cerviño, M., Martins, L. P., Leitherer, C., Hauschildt, P. H., 2005, MNRAS, 357, 945


\bibitem[]{} Erdelyi-Mendes, M., Barbuy, B., 1989, A\&AS, 80, 229
\bibitem[]{} Edvardsson, B., Andersen, J., Gustafsson, B., Lambert, D. L., Nissen, P. E., Tomkin, J., 1993, A\&A, 275, 101


\bibitem[]{} Fioc, K., Rocca-Volmerange, B., 1997, A\&A, 326, 950
\bibitem[]{} Fluks, M. A., Plez, B., The, P. S., de Winter, D., Westerlund, B. E., Steenman, H. C., 1994, A\&AS, 105, 311
\bibitem[]{} Franchini, M., Morossi, C., Di Marcantonio, P., Malagnini, M. L., Chavez, M., Rodriguez-Merino, L., 2004, ApJ, 601, 485
\bibitem[]{} Fuhrmann, K., Axer, M., Gehren, T. 1993, A\&A, 271, 451

\bibitem[]{} Girard, P., Soubiran, C., 2005, Proceedings of the Gaia
Symposium "The Three-Dimensional Universe with Gaia" (ESA SP-576). Held
at the Observatoire de Paris-Meudon, 4-7 October 2004. Editors: C.
Turon, K.S. O'Flaherty, M.A.C. Perryman, p. 169
\bibitem[]{} Goldman, A., Shoenfeld, W. G., Goorvitch, D., Chackerian C. Jr., Dothe, H., M\'elen, F., Abrams, M. C., Selby, J. E. A. 1998, JQSRT, 59, 453
\bibitem[]{} Gray, D., 1976, {\it The observation and analysis of stellar photospheres}, New York: Wiley-Interscience.
\bibitem[]{} Grevesse, N., Sauval, A. J., 1998, Space Science Reviews, v. 85, Issue 1/2, p. 161
\bibitem[]{} Goorvitch, D., 1994, ApJS, 95, 535
\bibitem[]{} Gunn, J.E., Stryker, L.L., 1983, ApJS, 52, 121


\bibitem[]{} Heavens, A., Panter, B., Jimenez, R., Dunlop, J., 2004,
Nature, 428, 625
\bibitem[]{} Hinkle, K., Wallace, L., Valenti, J., Harmer, D., 2000, {\it Visible and Near Infrared Atlas of the Arcturus Spectrum 3727-9300 A}, ed. Kenneth Hinkle, Lloyd Wallace, Jeff Valenti, and Dianne Harmer, San Francisco: ASP
\bibitem[]{} Holweger, H., M\"uller, E. 1974, Solar Phys., 39, 19 (HM74)
\bibitem[]{} Huber, K.P., Herzberg, G., 1979, {\it Constants of Diatomic Molecules}, van Nostrand Reinhold, New York         Spectra of Diatomic Molecules, Am.  Elsevier Pub.  Co.



\bibitem[]{} Jacoby, G.H., Hunter, G.A., Christian, C.A., 1984, ApJ, 419, 592
\bibitem[]{} Jimenez, R., MacDonald, J., Dunlop, J., Padoan, P., Peacock, J., 2004, MNRAS, 349, 240
\bibitem[]{} Jones, L.A. 1999, PhD Thesis, University of North Carolina, 
Chapel Hill
\bibitem[]{} Jorgensen, U. G., 1994, A\&A, 284, 179


\bibitem[]{} Katz, D.; The Rvs Team, 2004,  Semaine de l'Astrophysique Francaise, meeting held in Paris, France,Eds.: F. Combes, D. Barret, T. Contini, F. Meynadier and L. Pagani, EdP-Sciences, Conference Series 
\bibitem[]{} Kupka, F., Piskunov, N., Ryabchikova, T. A., Stempels, H. C., Weiss, W. W., 1999, A\&AS, 138, 119
\bibitem[]{} Kurucz, R.L., 1993, CD-ROM 13, 14, 18, 23
\bibitem[]{} Kurucz, R.L., Furenlid, I., Brault, J., 1984, {\it Solar flux atlas from 296 to 1300 nm}, National Solar Observatory Atlas, Sunspot, New Mexico: National Solar Observatory.


\bibitem[]{} Le Borgne, J.-F., Bruzual, G., Pelló, R., Lançon, A., Rocca-Volmerange, B., Sanahuja, B., Schaerer, D., Soubiran, C., Vílchez-Gómez, R., 2003, A\&A, 402, 433
\bibitem[]{} Leitherer, C., Schaerer, J., Goldader, J., Delgado, R., Robert, C. et al., 1999, ApJS, 123, 3
\bibitem[]{} Lejeune, Th., Cuisinier, F., Buser, R., 1997, A\&AS, 125, 229
\bibitem[]{} Lejeune, Th., Cuisinier, F., Buser, R., 1998, A\&AS, 130, 65
\bibitem[]{} Luque, J.,  Crosley, D. R. 1999, SRI International Report MP 99-009


\bibitem[]{} Maraston, C., 2005, submitted to MNRAS, astro-ph/0410207
\bibitem[]{} McWilliam, A., 1997, ARA\&A, 35, 503
\bibitem[]{} McWilliam, A. 1998, AJ, 115, 1640
\bibitem[]{} McWilliam, A., Rich, R. M., 1994, ApJS, 91, 749
\bibitem[]{} Martins, L.; Gonzalez Delgado, R. M.; Leitherer, C., Cervi\~no, M., Hauschildt, P., 2005, MNRAS 358, 49
\bibitem[]{} McWilliam, A., Rich, R. M., 1994, ApJS, 91, 749
\bibitem[]{} McWilliam, A., Rich, R. M., 2004, {\it Origin and Evolution of the Elements}, from the Carnegie Observatories Centennial Symposia. Carnegie Observatories Astrophysics Series. Edited by A. McWilliam and M. Rauch. Pasadena: Carnegie Observatories

\bibitem[]{} Mel\'endez, J., Barbuy, B., 1999, ApJS, 124, 527
\bibitem[]{} Mel\'endez, J., Barbuy, B., 2002, ApJ, 575, 474
\bibitem[]{} Mel\'endez, J., Barbuy, B. \&  Spite, F. 2001, ApJ, 556, 858
\bibitem[]{} Mel\'endez, J., Barbuy, B., Bica, E., Zoccali, M., Ortolani, S., Renzini, A., Hill, V., 2003, A\&A, 411, 417
\bibitem[]{} Mendes de Oliveira, C., Coelho, P., Gonzalez, J.J.,
  Barbuy, B., 2005, AJ, 130, 55
\bibitem[]{} Munari, U., Castelli, F., Zwitter, T., 2005, A\&A, accepted (astro-ph/0502047)  
\bibitem[]{} Murphy, T., Meiksin, A., 2004, MNRAS, 351, 1430



\bibitem[]{} O'Mara, B., 1976, MNRAS, 177, 551

\bibitem[]{} Panter, B., Heavens, A., Jimenez, R., 2004, MNRAS, 355, 764
\bibitem[]{} Pickles, A.J. 1998, PASP, 110, 863
\bibitem[]{} Plez, B., 1998, A\&A, 337, 495
\bibitem[]{} Plez, B., Brett, J.M., Nordlund, 1992, A\&A, 256, 551
\bibitem[]{} Phillips, J. G., Davis, S. P., 1968, {\it Berkeley Analyses of Molecular Spectra}, Berkeley: University of California Press
\bibitem[]{} Phillips, J.G., Davis, S.P., Lindgren, B., Balfour, W.J. 1987, ApJS, 65, 721
\bibitem[]{} Pomp\'eia, L., Barbuy, B., Grenon, M., 2003, ApJ, 592, 1173
\bibitem[]{} Praderie, F., 1967, Annales d'Astrophysique, Vol. 30, p.31
\bibitem[]{} Pradhan, A.D., Partridge, H., Bauschlicher, C.W.J.  1994,
  J. Chem. Phys., 101, 3857
\bibitem[]{} Prochaska, L.C., Rose, J.A. \& Schiavon, R.P., 2005, AJ,
submitted
\bibitem[]{} Proctor, R. N., Forbes, D. A., Beasley, M. A., 2004, MNRAS 355, 1327


\bibitem[]{} Reader, J., Wiese, W. L., Martin, W. C., Musgrove, A., Fuhr, J. R., 2002, NASA Laboratory Astrophysics Workshop, held May 1-3 2002 at NASA Ames Research Center, Moffett Field, CA 94035-1000. Publisher: NASA. Edited by Farid Salama. Reference Conference Proceedings: NASA/CP-2002-21186, p. 80.
\bibitem[]{} Rodriguez-Merino, L.H., Chavez, M., Bertone, E., Buzzoni, A 2005, ApJ, 626, 411
\bibitem[]{} Ryan, S. G., 1998, A\&A, 331, 1051


\bibitem[]{} Schiavon, R.P., Barbuy, B., Singh, P.D.: 1997, ApJ, 484, 499
\bibitem[]{} Schiavon, R.P., Barbuy, B., 1999, ApJ, 510, 934
\bibitem[]{} Schiavon, R.P., 2005, ApJS, submitted
\bibitem[]{} Schulz, R.H., Armentrout, P.B. 1991, J.Chem.Phys., 94, 2262
\bibitem[]{} Schulz, J., Fritze-v. Alvensleben, U,, Möller, C. S., Fricke, K. J., 2002, A\&A, 392, 971
\bibitem[]{} Sneden, C., McWilliam, A., Preston, G. W., Cowan, J. J., Burris, D. L., Armosky, B. J., 1996, ApJ, 467, 819
\bibitem[]{} Spielfiedel, A., Feautrier, N., Chambaud, G., Levy, B., 1991, Journal of Physics B: Atomic, Molecular, and Optical Physics, Volume 24, Issue 22, p. 4711
\bibitem[]{} Spite, M.: 1967, Ann. d'Astroph., 30, 211
\bibitem[]{} Steffen, M., 1985, A\&AS, 59, 403
\bibitem[]{} Steinmetz, M., 2003, ASPC, 298, 381


\bibitem[]{} Tantalo, R., Chiosi, C., 2004, MNRAS 353, 917
\bibitem[]{} Thomas, D., Maraston, C., Bender, R., 2003a, MNRAS, 339, 897
\bibitem[]{} Thomas, D., Maraston, C., Bender, R., 2003b, MNRAS, 343, 279
\bibitem[]{} Timmes, F. X., Woosley, S. E., Weaver, Thomas A., 1995, ApJS, 98, 617
\bibitem[]{} Trager, S.C., Worthey, G., Faber, S.M., Burstein, D., Gonz\'alez, J.J. 1998, ApJS, 116, 1
\bibitem[]{} Tsuji, T. 1973, A\&A, 23, 411

\bibitem[]{} Unsold, A., 1955, {\it Physik der Sternatmospharen, MIT besonderer Berucksichtigung der Sonne}, Berlin: Springer.


\bibitem[]{} Valdes, F., Gupta, R., Rose, J., Singh, H., Bell, D., 2004, ApJS, 152, 251
\bibitem[]{} Valenti, J.A. \& Fischer, D.A. 2005, ApJS, 159, 141
\bibitem[]{} van 't Veer-Menneret, C., Megessier, C. 1996, A\&A, 309, 879
\bibitem[]{} Vazdekis, A., 1999, ApJ 513, 224


\bibitem[]{} Westera, P., Lejeune, T., Buser, R., Cuisinier, F., Bruzual, G., 2002, A\&A 381, 524
\bibitem[]{} Willemsen, P.G, Hilker M., Kayser, A., Bailer-Jones,
C.A.L., 2005, A\&A, in press, astro-ph:0503345
\bibitem[]{} Worthey, G., Faber, S. M., Gonzalez, J. J., 1992, ApJ, 398, 69
\bibitem[]{} Worthey, G., Faber S.M., Gonz\'alez, J.J., Burstein, D., 1994,
ApJS, 94, 687


\bibitem[]{} Zoccali, M., Barbuy, B., Hill, V., Ortolani, S., Renzini, A., Bica, E., Momany, Y., Pasquini, L., Minniti, D., Rich, R. M., 2004, A\&A, 423, 507
\bibitem[]{} Zwitter, T., Castelli, F., Munari, U., 2004, A\&A, 417, 1055

\end{thebibliography}
\end{document}